\documentclass[sigconf,screen,authorversion,nonacm]{acmart}

\usepackage{enumitem}
\usepackage{subcaption}
\usepackage{wrapfig}
\usepackage{colortbl}
\usepackage{graphicx}
\usepackage{acmart-taps}

\graphicspath{source/}

\definecolor{mycolor1}{RGB}{204,121,167}
\definecolor{mycolor2}{RGB}{213,94,0}
\definecolor{mycolor3}{RGB}{0,114,178}
\definecolor{mycolor4}{RGB}{240,228,66}
\definecolor{mycolor5}{RGB}{0,158,115}

\AtBeginDocument{  \providecommand\BibTeX{{    \normalfont B\kern-0.5em{\scshape i\kern-0.25em b}\kern-0.8em\TeX}}}

\newcommand{\httpsurl}[1]{\href{https://#1}{\nolinkurl{#1}}}

\begin{document}
\emergencystretch 3em

\title{Queer In AI: A Case Study in Community-Led Participatory AI}

\small
\author{Organizers Of QueerInAI, Anaelia Ovalle, Arjun Subramonian, Ashwin Singh, Claas Voelcker, Danica J. Sutherland, Davide Locatelli, Eva Breznik, Filip Klubi\v{c}ka, Hang Yuan, Hetvi J, Huan Zhang, Jaidev Shriram, Kruno Lehman, Luca Soldaini, Maarten Sap, Marc Peter Deisenroth, Maria Leonor Pacheco, Maria Ryskina, Martin Mundt, Milind Agarwal, Nyx McLean, Pan Xu, A Pranav, Raj Korpan, Ruchira Ray, Sarah Mathew, Sarthak Arora, ST John, Tanvi Anand, Vishakha Agrawal, William Agnew, Yanan Long, Zijie J. Wang, Zeerak Talat, Avijit Ghosh, Nathaniel Dennler, Michael Noseworthy, Sharvani Jha, Emi Baylor, Aditya Joshi, Natalia Y. Bilenko, Andrew McNamara, Raphael Gontijo-Lopes, Alex Markham, Evyn D\v{o}ng, Jackie Kay, Manu Saraswat, Nikhil Vytla, Luke Stark}
\affiliation{%
  \institution{Queer in AI}
  \country{}
}

\renewcommand{\shortauthors}{Organizers of QueerInAI, et al.}

\begin{abstract}
Queerness and queer people face an uncertain future in the face of ever more widely deployed and invasive artificial intelligence (AI). These technologies have caused numerous harms to queer people, including privacy violations, censoring and downranking queer content, exposing queer people and spaces to harassment by making them hypervisible, deadnaming and outing queer people. More broadly, they have violated core tenets of queerness by classifying and controlling queer identities. In response to this, the queer community in AI has organized Queer in AI, a global, decentralized, volunteer-run grassroots organization that employs intersectional and community-led participatory design to build an inclusive and equitable AI future. 
In this paper, we present Queer in AI as a case study for community-led participatory design in AI. We examine how participatory design and intersectional tenets started and shaped this community's programs over the years. We discuss different challenges that emerged in the process, look at ways this organization has fallen short of operationalizing participatory and intersectional principles, and then assess the organization's impact. Queer in AI provides important lessons and insights for practitioners and theorists of participatory methods broadly through its rejection of hierarchy in favor of decentralization, success at building aid and programs by and for the queer community, and effort to change actors and institutions outside of the queer community. Finally, we theorize how communities like Queer in AI contribute to the participatory design in AI more broadly by fostering cultures of participation in AI, welcoming and empowering marginalized participants, critiquing poor or exploitative participatory practices, and bringing participation to institutions outside of individual research projects. Queer in AI's work serves as a case study of grassroots activism and participatory methods within AI, demonstrating the potential of community-led participatory methods and intersectional praxis, while also providing challenges, case studies, and nuanced insights to researchers developing and using participatory methods. 
\end{abstract}

\maketitle

\begin{figure*}[!ht]
\centering
\includegraphics[scale=0.38]{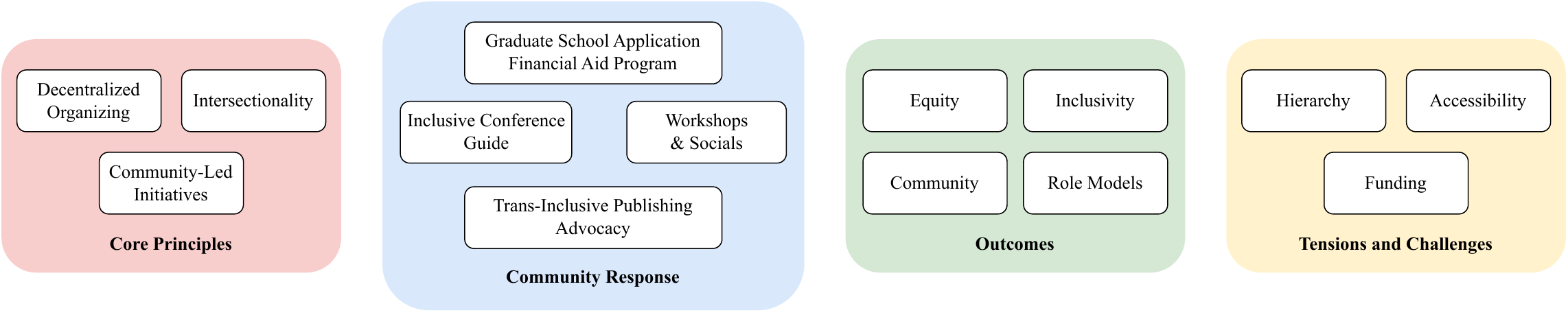}
\caption{Overview of Queer in AI's core principles, community responses, programming outcomes, and tensions and challenges.}
\label{fig:qinai_flowchart}
\end{figure*}

\section{Introduction}

Artificial intelligence (AI) has seen enormous developments in recent years, such as substantial advances in protein modeling, drug discovery, weather prediction, and personalized medicine~\cite{ravuri2021skilful,AlphaFold2021,vamathevan2019applications}. The ubiquity of unregulated AI within socio-technical systems, however, often produces discriminatory outcomes and harms marginalized communities globally~\cite{khari2022wrongful, bender2021dangers, birhane2021multimodal}.
For queer people in particular, machine learning models learn brittle, toxic representations that cause representational and allocational harms, from misgendering to healthcare discrimination~\cite{keyes2018misgendering, tomasev2021fairness, dev2021harms, dodge2021documenting}. Identifying and mitigating harmful outcomes has led to the development of computational and socio-technical methods for achieving fairness~\cite{mehrabi2021, blodgett-etal-2020-language, costanza2018design}, including automatic evaluation and unfairness mitigation techniques~\cite{Dwork2011FairnessTA, Bolukbasi2016ManIT, mehrabi2021}. 
While such approaches have the potential to mitigate harms for queer people in domains like fighting online abuse, health, and employment~\cite{tomasev2021fairness}, computational techniques generally encode narrow conceptualizations of fairness where queer identities are assumed to be known, observable, measurable, discrete, and static \cite{lu2022subverting}. By locating the source of unfairness in individuals or in specific design decisions~\cite{weinberg2022rethinking}, computational approaches to fairness can reinforce existing power relations~\cite{d2020data, kalluri2021don},
including marginalized communities only in predatory ways~\cite{beyondfairness2021} or as ``ethics washing''~\cite{sloane22participation}
(cf.\ Appendix~\ref{app:fairness_limitations} for an extended critique of computational approaches to fairness).

Participatory methods address some of these limitations. Involving users as co-designers holds great potential for dismantling power relations and empowering marginalized communities that are disproportionately impacted by AI~\cite{abeba22power, suresh22pml,kormilitzin2023participatory}. Reflexivity in participatory methods encourages transparency during the design process itself, as opposed to a detrimental ``innovate first, fix later'' approach to building trustworthy AI~\cite{floridi2019establishing}.
By establishing the value-laden nature of technologies, it can prevent personal biases, beliefs and values from seeping into AI systems unexamined. 

Unfortunately, there are many challenges to incorporating participatory approaches across top-down structures, such as corporations that operate within capitalism. Popular modes of participation within AI suffer from extractive and exploitative forms of community involvement or ``participation washing''~\cite{sloane22participation}. For example, a recent report~\cite{time-chatgpt} sheds light on how OpenAI used exploitative labor practices to make ChatGPT less toxic, subjecting Kenyan workers to psychologically distressing content\footnote{This content included examples of sexual abuse, hate speech, violence, murder, child abuse, rape, animal abuse, torture and self-harm.} without sufficient provision for mental health support. \citet{gray2019ghost} also uncovers many exploitative labor practices performed by minorities to power AI systems.

More fundamentally, we question whether marginalized communities should engage in designing with the creators of harmful AI systems that prioritize profit over their safety.
Even in projects where communities are involved, engagement is too often limited in scope and time. Contrary to participation being controlled by the corporations and states the design and own AI, we argue in the favor of shifting power towards marginalized groups and centering their experiences. 
We call for a culture of participation in AI to address this, one that enables deep and long-term participation in AI research, institutions, and practices. 

Over the years, the AI community has witnessed several community-led efforts from marginalized communities,
each tackling issues of inequality that arise along various axes of marginalization;
these include Black in AI~\cite{blackinai}, 
LatinX in AI~\cite{latinxinai}, Women in Machine Learning~\cite{wiml}, Masakhane~\cite{masakhane}, Widening NLP~\cite{winlp-org}, Diversity in AI~\cite{diverseinai}, Indigenous in AI~\cite{indigenousinai}, Queer in HCI~\cite{devito2020queer}, the Indigenous Protocol and AI Working Group~\cite{lewis2020indigenous}, the Deep Learning Indaba~\cite{dlindaba}, Khipu~\cite{khipu}, North Africans in ML~\cite{northafricans}, \{Dis\}Ability in AI~\cite{disai}, Te Hiku Media~\cite{finn2022developing}, and Muslims in ML~\cite{muslimsinml}.
These organizations have worked in AI ethics, advocated against AI harms, provided longstanding venues and visibility for AI ethics research within major ML and NLP conferences, resolved inclusion issues with those venues, and developed community-led datasets, models, and other technology. Most importantly, they have advanced participation by marginalized communities in AI research and development at large, nurturing countless researchers and practitioners with community, mentorship, financial aid, and innumerable other forms of help with the many barriers marginalized people face in AI. These groups have made AI much more diverse, and strengthened the voices of marginalized people within AI.

In this work, we argue that AI ethicists who value participatory methods as a means for making ethical AI should engage with participatory and community-lead AI ethics organizations, and study their organizational, strategic, and administrative work through which they are advancing participation and building cultures of participation. This often difficult process involves navigating the complexities of combining inquiry with praxis, and sheds light on  differences between participatory approaches.

To this end, we offer a case study analyzing Queer in AI, a grassroots organization that aims to raise awareness of queer issues in AI/ML, foster a community of queer researchers and celebrate the work of queer scientists. Operating primarily as an online community over Slack, the organization runs various programs and initiatives towards fulfilling its mission. We analyze and critique its principles, methodology, initiatives, and impact over the years as a case study of community-led participatory methods in AI.

Our key contributions are:

\begin{itemize}
    \item We document salient forms of marginalization and oppression that particularly affect queer people (\S\ref{sec:marginalization}).
    \item We present the organizing principles and programs of Queer in AI (\S\ref{sec:principles}), including how they started, major changes, and qualitative and quantitative analyses of impacts (\S\ref{sec:initiatives}).
    \item We analyze challenges and shortcomings of Queer in AI (\S\ref{sec:discussion}).
    \item We present an argument for conducting more and valuing AI ethics research that combines inquiry and praxis (\S\ref{sec:conclusion}).
\end{itemize}

\textit{Positionality Statement \quad}
Most authors of this paper are formally trained as computer scientists, with some also having training in gender theory or related fields.
All authors have informal training in queer studies through activism and advocacy. Our backgrounds influence this work's design, decisions, and development. We do our best to position our work in a global context, with authors from Asia, Europe, South Africa, South America, and North America.

\section{Marginalization of queer people in STEM and AI}\label{sec:marginalization}
Hegemonic forms of AI focus on classifying complex people and situations into narrow categories at the cost of context,
and 
are often built to support surveillance, prediction, and control
-- designs which are fundamentally incompatible with queer identities rooted in the freedom of being~\cite{keyes2019counting}.
The framing and use of common AI systems that interact with gender are thus often problematic, and inherently cisnormative and heteronormative, so that even well-meaning, purportedly inclusive AI projects are prone to ``designing out'' certain queer lives~\cite{guyan2022fixing}.
Documented harms across various AI applications are numerous, and sometimes life-threatening.
These include physiognomic and phrenologic applications such as computer vision to (falsely) infer gender and sexuality~\cite{newclothes, stark2021physiognomic, keyes2018misgendering, scheuerman2019computers, scheuerman2021auto, long2021agr, Keyes2021TruthFT}.
AI-enabled surveillance systems, in conjunction with surveillance of online spaces such as dating apps by states, corporations, and even individuals have outed queer people, compromising their privacy and safety~\cite{ceres_2022, priestouted, olympiansouted, grindrpolice}.
Online spaces, especially social media platforms, have insufficient and poorly explained privacy and security tools, requiring community education and adaptation to meet the needs of queer people~\cite{geengqueer, devito2018too, pinter2021entering}.
Their moderation enables widespread censorship of queer words and identities~\cite{salty, noaccess, wiredlbnoobw, elefante2021lips}, while also subjecting queer communities to disproportionate online harassment and hate speech~\cite{powell2020digital, treebridge2021dogwhistle}. 
Some of these harms can be traced to large language models (LLMs) trained on datasets containing hate speech and censored queer words, leading search systems to avoid queer content and content moderation systems to more often tag it as suspect~\cite{gomes2019drag, dodge2021documenting}. LLMs also overwhelmingly fail to account for non-binary genders and pronouns, contributing to erasure of these identities~\cite{dev2021harms, cao-daume-iii-2020-toward}.

In the US, queer people are (at least) 20\% less represented in STEM than in the national population, and experience higher levels of ``career limitations, harassment, and professional devaluation''~\cite{cech2021systemic}. Consequently, queer scientists often face ``systematically more negative workplace experiences than their non-LGBT colleagues'' \cite{Cech2017QueerIS}, and ``leave STEM at an alarming rate'' \cite{Freeman2020MeasuringAR}. The exclusion of queer people from science comes with significant consequences, both for queer scientists and queer people further marginalized by fields that do not understand or care about them.  The medical profession's response to the HIV/AIDS crisis was fatally slow until pressured by heroic activism~\cite{schulman2021let}; a medical field that had included and empowered queer people may have saved many queer lives. Similarly, the American Psychiatric Association classified homosexuality as a mental illness until 1973, greatly contributing to the stigmatization of queer people around the world, until queer activists pressured the group for change~\cite{drescher2015out}. Recent initiatives have inverted this dynamics, centering queer communities in descisions about mental healthcare~\cite{kormilitzin2023participatory}.

One hurdle in understanding the marginalization of LGBTQIA+ people in STEM is a lack of demographic data on sexual orientation and gender identity \cite{Freeman2020MeasuringAR}. 
The US's National Science Foundation has delayed the collection of such data for years, despite the urging of queer scientists \cite{nsf-data-collection}.
Taking matters into its own hands, Queer in AI administers an annual survey of its global community to uncover the demographics and challenges faced by queer researchers in AI (discussed in detail in Appendix~\ref{app:surveys}).
In Queer in AI's 2021-22 community survey ($N = 252$), 74\% of members reported a lack of role models and 77\% reported a lack of community as obstacles in their journey of becoming an AI practitioner. 

There is a dire lack of studies and data on queer scientists' experiences in the Global South, where colonial histories have led to the criminalization of queerness \cite{brazil-qstem, SA-qstem, india-qstem}. Queer in AI organizers from Turkey, Colombia, and India have shared that much queer activism in these countries focuses on survival and gaining basic human rights, recognition and respect in society, amid high levels of discrimination, violence, and psychological distress \cite{lgbt-colombia-2019}. They perceive being out and working towards queer visibility in STEM fields to be beyond luxuries, especially given the dominant (cisnormative, heteronormative) view that identity and profession should be ``kept separate.'' Barriers to acceptance are only amplified for queer individuals also marginalized on intersecting axes like class or caste.

\section{Core Principles of Queer in AI}\label{sec:principles}
Three governing principles drive Queer in AI’s mission to raise awareness of queer issues in AI and foster a community of queer researchers: (i) decentralized organizing, (ii) intersectionality, and (iii) community-led initiatives. Overall, Queer in AI's decentralized operations allow for swift community-led initiatives towards its mission (\S\ref{sec:decentralized}), which center on intersectionality as critical inquiry and praxis (\S\ref{sec:intersectionality}). In doing so, it acknowledges and continuously works to account for ``the complexities of multiple, competing, fluid, and intersecting identities'' \cite{gringeri2010mapping}. Queer in AI's primary approach consists of including people with diverse lived experiences in participatory schemes (\S\ref{sec:community}).

\subsection{Participation and Decentralization} 
\label{sec:decentralized}

For its first two years, Queer in AI had a hierarchical structure, with a president and officers. However, organizing and governance of grassroots communities, and especially queer communities, presents unique challenges. Queer people are incredibly diverse, and choosing one or even a group of queer people to represent the community as a whole is reductive and impossible. This is also difficult for the organizers, with high-profile queer activists and organizers frequently facing targeted harassment campaigns, and Queer in AI organizers frequently reporting lack of time, external support, or recognition for volunteering (Figure~\ref{fig:organizer_challenges}). Queer in AI thus adopted a decentralized organizing structure, to encourage broad participation. Queer in AI minimizes distinctions between organizers and members to encourage the entire community to participate in organizing. Most volunteer coordination occurs in the same Slack channel as is used for community discussion, calls for help or feedback on programs mixed with memes, introductions, personal news, and discussions of travel or pets.
Of the 49 active Slack channels only 4, where personally identifiable information is discussed, are not public.
Openness and embedding in the community increase transparency and accountability: any community member can view organizing discussions and join in, with no more barrier to entry than joining a Slack channel.
It also helps provide the connection and joy for which 75\% of its organizers joined Queer in AI (Figure~\ref{fig:organizer_motivation}).
Fluidity between member and organizer also makes it easier for community members' areas and levels of engagement to ebb and flow over time without losing their connection to the community. 

\subsection{Participation and Intersectionality}
\label{sec:intersectionality}
\begin{figure*}
    \centering
    \caption{Country of origin of the respondents to the Queer in AI's 2021--2022 demographic survey.}
    \includegraphics[width=0.7\linewidth]{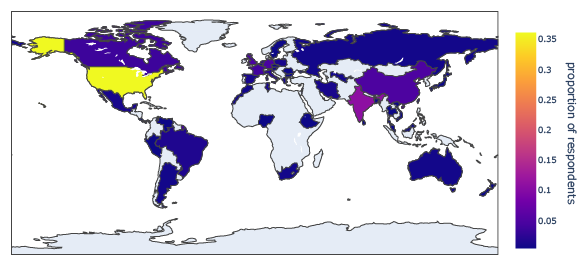}
    \label{fig:country_heatmap}
\end{figure*}
\begin{table*}
\centering
\footnotesize
\caption{Self-reported ethnicity, gender, and sexual orientation of the respondents to the Queer in AI's 2021--2022 demographic survey. Write-in responses were aggregated by a team of Queer in AI organizers, with some falling into multiple categories (see \S\ref{ssec:aggregation}). ``Unaggregated'' refers to responses that could not be adequately described with any subset of other categories; however, responses in this group may overlap with the remaining categories. For options with fewer than 4 responses, exact values are omitted for privacy.}
\label{tab:member_identities}
\begin{tabular}{lrllrllr}
\toprule
\multicolumn{2}{c}{\textbf{Ethnicity}} & & \multicolumn{2}{c}{\textbf{Gender}} & & \multicolumn{2}{c}{\textbf{Sexual Orientation}} \\
\midrule
    Caucasian & 127	& & Man & 108 & & Queer & 90 \\
    South Asian & 34 & & Woman & 95 & & Gay & 89 \\
    East Asian & 17 & & Non-binary & 61 & & Bisexual & 87 \\
    Black/African/African-American & 13 & & Genderqueer & 29 & & Pansexual & 42 \\
    Latinx & 13	& & Gender non-conforming & 22 & & Lesbian & 30 \\ 
    Mixed & 12 & & Genderfluid & 19 & & Asexual & 26 \\
    Jewish & 8 & & Agender & 17 & & Unaggregated & 29 \\
    Middle Eastern & 8	& & Questioning & 16 & & & \\
    Southeast Asian & 6 & & Unaggregated & 16 & & & \\
    West Asian & $\leq$ 3 \\
    Central Asian & $\leq$ 3 \\
    Hispanic & $\leq$ 3 \\
    Unaggregated & 6 \\
\bottomrule
\end{tabular}
\end{table*}

Over five years, Queer in AI’s community has grown to about 870 members, geographically distributed across more than 47 countries (cf.\ Figure \ref{fig:country_heatmap}). The community members have diverse identities across axes such as ethnicity, gender, class, disability, and caste. About 20.3\% of respondents identified as transgender, and 34.4\% identified as non-cisgender; 34.9\% identified as Black, Latinx, indigenous or a person of color; less than 2\% identified as intersex.
Membership spans academia and industry, with about 16\% of members pursuing an undergraduate degree, 21\% in an industry role, and 64\% in academia, all with varying degrees of seniority (cf.\ Appendix~\ref{app:surveys} for additional details of community demographics).
As a result, Queer in AI helps naturally bridge otherwise insular aisles of power and social contexts.

As the queer community consistently experiences discrimination, stigmatization, and inequity \citep{meyer2015violence, casey2019discrimination}, Queer in AI uses the lens of intersectionality as a means of critical inquiry to identify how interlocking forms of oppression, such as racism and sexism, co-construct and exacerbate social and structural disparities \cite{collins2020intersectionality}.
To proactively dismantle injustices, Queer in AI centers the experiences of its members so that active participation in the Queer in AI community results in the co-creation of initiatives, which reflect of tackling such barriers, including economic (\S\ref{sec:funding}), educational (\S\ref{sec:grad-app}), and social (\S\ref{sub:workshops}) ones. By prioritizing fighting intersectional oppression, Queer in AI attempts to empower its most marginalized members to shape and control its programming, addressing key challenges of participatory design such as the exclusion of marginalized people from participation \citep{katell2020toward}, community power-sharing \citep{collins2018community} and the co-formation of knowledge \citep{ferree2016discursive}. In doing so, Queer in AI works towards a system of resistant knowledge firmly grounded in intersectionality's critical praxis \cite[Chs.\ 3 \& 4]{collins2019intersectionality}.

\subsection{Participation and Community Leadership} 
\label{sec:community}

\subsubsection{Research}Various forms of community-engaged research guide the dissemination of knowledge both within and outside of Queer in AI and exist across a continuum, from community-informed to community-involved to community-led. Community-informed research consists of researchers inviting the community to incorporate lived experience to guide research questions, data collection, or data interpretation \citep{hacker-taylor}.
Towards more community-involved research, community members may be more involved in decision-making processes and research planning \citep{russell2008acquire, hacker-taylor}. At the highest level of engagement, community-driven approaches such as community-based participatory action research (PAR) centers shared collaborative decision-making between researchers and community members across research design, knowledge creation, intervention development, and policy-making \citep{community-based-participatory-action-research, maiter2008reciprocity, cooke2001participation}.
In practice, entities outside of the organization may partner with Queer in AI community members to form relationships designed to help objectives oriented towards investigating and supporting ``the pursuit of answers to the questions of their daily struggle and survival'' \citep{tandon1988social}. Individuals are often members of both other entities as well as of Queer in AI so that members may operate from the role of an external entity (e.g.\ researcher from a company) and at various depths of community engagement.
The resulting knowledge production is such that is ``by the people, for the people'' in which research is not only seen as a process to create knowledge but to also educate and mobilize for action \citep{cooke2001participation, green2003appendix}. By ``putting community first'', the distinction between participant and researcher is removed. Community-based participatory action research thus also serves as a decolonizing epistemological framework which inherently interrogates power and privilege \citep{fine2006intimate}.

\subsubsection{Response \& resilience} Within Queer in AI, community resilience operates across dimensions including but not limited to the social, political, and economic. Advocacy efforts operate across domains, tasks, resources, and activities within the organization \citep{kreps1994organizing}. Resources
and activities
are structural means towards tasks
and domains
that reflect the Queer in AI mission. Specifically, resources and activities are dedicated to raising awareness of queer issues in AI/ML. Financial, educational, and social avenues are created within the organization as a form of creating resilience and advocacy in the face of oppressive sociotechnical barriers.
Operating across 47 countries, Queer in AI  primarily organizes through Slack, Zoom, a dedicated mailing list, and social media platforms. Doing so makes room for rapid and adaptive situational awareness within the online community \cite{starbird2011voluntweeters}. Besides the ``internal'' milieu of an organization, Queer in AI is responsive to events in both reactive and proactive forms. Digital volunteer efforts emerge as self-organizing responses to external factors \citep{dynes1970organized, cobb2014designing}. This work further details examples of how responses to acute external factors and larger efforts against oppression manifest as Queer in AI initiatives.

\section{Queer in AI Initiatives} \label{sec:initiatives}
The structure of Queer in AI is decentralized and includes volunteers, core organizers (extensive organizing experience with Queer in AI) and a diversity, equity and inclusion admin (DEIA, a core organizer who has a more active role in administrative duties). Most of Queer in AI’s communication is mediated by its Slack workspace. 

A key aspect of Queer in AI’s organizing lies in the transparency of its operations and associated information exchanges, which predominantly take place over public Slack channels. There are only four private channels on the workspace, which exist to preserve privacy while facilitating discussions around personally identifiable information. The workspace has included the exchange of over 133,000 messages (including individuals' one-to-one private messages), of which over 25,000 have been sent in public channels, accounting for the majority (57\%) of total views. This transparency, in conjunction with regular updates and outreach on Slack, keeps community members involved in ongoing events and initiatives. Many of Queer in AI’s initiatives have emerged from conversations and threads on public channels about discriminatory experiences with different institutions.
For example, discussion around exclusionary gender collection practices on conference registration forms led to the creation of an inclusive conference guide (covered in more detail in \S\ref{sub:inclusive_conf_guide}) and substantial improvements to relevant conferences' practices.
Similarly, significant advocacy against deadnaming in citations and conference proceedings (\S\ref{sub:namechange}) began from discourse on public channels. Thus, as a space, Queer in AI’s Slack is effective at mobilizing community-led initiatives through decentralized organizing. Moreover, the emergence of these initiatives from diverse yet intersecting shared queer experiences grounds them in global contexts of social inequality and injustice.
For instance, Queer in AI’s graduate school application financial aid program (\S\ref{sec:grad-app}) and workshops and socials (\S\ref{sub:workshops}) target several particular challenges rooted in non-Western contexts, centering otherwise-marginalized experiences.
The organizational and volunteer work that constitutes the administration of all these initiatives is thus deeply intersectional.

We now examine four major initiatives in detail;
Appendix~\ref{sec:policy} further describes efforts in policy advocacy.

\subsection{Graduate School Application Financial Aid Program} 
\label{sec:grad-app}

\begin{table*}[ht]
    \centering
    \footnotesize
    \caption{The Queer in AI Graduate School Application Fee Aid Program budget and impact per academic year, in USD.\label{tab:grad_app_numbers}}  
    \vspace{-1em}
    \begin{tabular}{p{3.2cm}p{2.3cm}ccc}
        \toprule
         Academic year & Aid per applicant & No. aid recipients & Total aid & Budget\\
         \midrule
         2020/2021 & up to \$750 & 31 & \$16,689 & \$20,000 \\
         2021/2022 & up to \$1,250 & 81 & \$70,607 & \$73,768 \\
         2022/2023 \footnotesize{(at time of writing)} & up to \$1,250  & 48 & \$40,476  & \$41,711  \\
         \bottomrule
    \end{tabular}
\end{table*}

\begin{table*}[ht]
\centering
\footnotesize
\caption{Gender, sexual orientation, romantic orientation and continent of scholarship recipients who filled the optional feedback survey ($n = 46$ out of $N = 160$ total recipients). For options with fewer than 4 responses, exact values are omitted for privacy.}
\label{tab:sr_all}
\vspace{-1em}
\begin{tabular}{lrllrllrllr}
\toprule
\multicolumn{2}{c}{\textbf{Gender}} &  & \multicolumn{2}{c}{\textbf{Sexual Orientation}} &  & \multicolumn{2}{c}{\textbf{Romantic Orientation}} & & \multicolumn{2}{c}{\textbf{Continent}} \\
\midrule
    Woman & 20 &  & Gay & 18 &  & Homoromantic & 21 & & Asia & 19 \\
    Man & 18 &  & Queer & 16 &  & Biromantic & 13 & & North America & 14 \\
    Genderqueer & 7 &  & Bisexual & 12 &  & Demiromantic & 5 & & Africa & 5 \\
    Non-binary & 6 &  & Lesbian & 9 &  & Grayromatic & 5 & & Europe & $\leq$3 \\
    Gender non-conforming & 6 &  & Asexual & 4 &  & Alloromantic & $\leq$3 & & South America & $\leq$3 \\
    Agender & $\leq$3 &  & Pansexual & 4 &  & Aromantic & $\leq$3 & & & \\
    Genderfluid & $\leq$3 &  & Demisexual & $\leq$3 &  & Heteroromantic & $\leq$3 & & & \\
    Questioning & $\leq$3 &  & Questioning & $\leq$3 &  &  & & & & \\
\bottomrule
\end{tabular}
\end{table*}

Queer folks report a lack of community and queer role models due to the underrepresentation of senior queer folks in academia. Thus, supporting queer and low-income scholars financially helps bring more marginalized voices into STEM academia, creating more opportunities for participatory research and technology design. To address this, Queer in AI launched the Graduate School Application Fee Aid Program to improve queer representation and make graduate programs accessible.

\subsubsection{Financial challenges}

The costs for graduate school applications prevent many low-income and international scientists from accessing graduate programs, well before they can benefit from many of the fellowships and need-based scholarships intended to address exclusion.
This process is costly: between the application fees ($\sim$\$50--\$150 USD per program in North America and parts of Europe), costs of required tests (e.g.\ GRE), test results and transcript delivery fees, and test preparation expenses, one round of applications can easily amount to over \$1,000 USD. 
International applicants may be further required to pay for language proficiency tests (e.g.\ TOEFL), translation services, and third-party credential vetting. 
Although some schools offer fee waivers, they vary widely from school to school, are often very limited in applicability, and can require onerous documentation.

The majority of applicants apply to North American schools. 
This is likely caused by the cultural dominance of Anglo-American schools in the AI/ML space and the common practice of requiring extensive standardized tests and application fees at these schools.\footnote{While fees and standardized tests are the norms at many prominent institutions, there are examples of alternative paths, such as the ELLIS PhD Program, a European initiative for AI/ML PhD programs, which requires neither \cite{ellis-call}.} 
Standardized tests like the GRE claim to level the playing field for applicants, they
institute barriers to individuals from the Global South and reify colonialism under a veneer of fairness.
Additionally, fees make these exams wholly inaccessible to many in the Global South: the GRE costs three times the average monthly salary in Ethiopia \citep{blackinai-ac-program}.
Data collected from Queer in AI's surveys have been used to argue that departments should eliminate the GRE and application fees.

These financial challenges are particularly likely to be insurmountable for queer scientists, who may be cut off from familial financial support, might pay out of pocket for gender-affirming healthcare, and often incur additional expenses managing oppression and trauma.
Queer people thus suffer from increased student loan debt~\cite{studentloan} and high rates of housing insecurity~\cite{wilson2020homelessness}. 
A complete critique of the graduate application process and its socio-economical context is out of the scope of this paper.
Queer in AI believes it is nonetheless important to provide concrete aid right now to applicants faced with the current system.

\subsubsection{Mutual aid design}
The design of the aid program is decentralized, community-led constituting volunteers with a diverse range of experiences with graduate school admissions \cite{spade2020mutual}. 
This initiative keeps minimum barriers to receiving the aid by not seeking to decide who is ``deserving'' of aid, avoiding imposing excessive requirements for documenting eligibility and providing timely mentorship and help to the applicants for their submissions.
Although, the payment pipeline often disadvantages applicants from countries and territories where PayPal is not available or restrictions are imposed on receiving transfers from the US.

\subsubsection{Participatory learnings}
Each aid applicant is treated as a member of the community with a valuable perspective of their own -- the initiative actively seeks feedback from aid recipients and encourages them to volunteer in the future, which would both help improve the program and keep it sustainable. 
This feedback indicated that aid recipients' demographics were more diverse than Queer in AI’s organizing team (Table~\ref{tab:sr_all}), which helps Queer in AI recruit more diverse volunteers and community members by first directly, meaningfully helping them.
Also, the feedback survey illustrates widespread deficiencies in existing admissions fee waivers: 
such as lack of fee waivers (67\%), unable to produce adequate documentation (14\%) and the fear of outing themselves (10\%). 
This aid program allowed recipients to take admissions tests (56\%), avoid skipping essential expenses (54\%) and avoid skipping groceries or bills (40\%).
The vast majority of recipients reported the scholarship enabled them to apply to additional programs (around 6 on average).

\subsubsection{Critical Reflections}
The program operates with a tension between opening opportunities to marginalized people from all over the world and reinforcing the exclusionary practices of these powerful institutions. 
In addition to funding influential and rich academic institutions, the program also indirectly supports the standardized testing industry.
The limited amount of funds and barriers to sending the money internationally often pose challenges between the organizers and the aid recipients. 
In spite of that, Queer in AI believes that it is crucial to provide timely aid regardless of these barriers, even if doing so reinforces undesirable structures.

\subsection{Workshops and Socials}
\label{sub:workshops}

In STEM disciplines, conferences can be a hostile setting for minoritized groups~\cite{Richey2019GenderAS,yadav2020forgotten,mcmillon2021implementing}.
Queer in AI members in 2022 rated how welcome they felt attending AI conferences at 3.38 on average ($\mu_{1/2}$ = 3) on a five-point Likert scale (\S\ref{app:surveys}). Recognizing this need, Queer in AI has organized workshops and networking events since its very first informal meetup at NeurIPS 2017: as of submission, 13 workshops and 35 social events in total (Table~\ref{tab:events}), with a cumulative attendance of hundreds of participants.\footnote{An exact count could not be obtained: to maintain attendees' privacy, Queer in AI does not require signups for most events, and deletes names immediately after events when they are required.} These events provide an opportunity to connect and network with other queer scientists, spotlight work by members of Queer in AI, host talks on topics relevant to its members, and arrange panels where experts discuss topics at the intersection of AI, fairness, ethics, and the queer community. 
The following subsections cover how Queer in AI’s principles influence event planning and enable them to overcome challenges in the process.

\aptLtoX[type=html, graphics=no]{
\begin{table*}[t]
\caption{Workshop and events organized by Queer in AI in 2017--2022 across conferences in AI. Events marked with \textit{p} were held in person, \textit{v} indicates virtual-only events,  and \textit{h} refers to events that occurred in a ``hybrid'' format. \label{tab:events}}
\vspace{-1em}
\footnotesize 
\begin{tabular}{@{}ccccccc@{}}
\toprule
\textbf{Year} & 
    \textbf{2017} & 
    \textbf{2018} & 
    \textbf{2019} & 
    \textbf{2020} & 
    \textbf{2021} & 
    \textbf{2022} \\
\midrule
\textbf{Workshops} & 
    \textbf{-} & 
    \begingroup
    \renewcommand*{\arraystretch}{0.6}
    \begin{tabular}[c]{@{}c@{}}
        \text{1} \\ 
        {\tiny NeurIPS$^\textit{p}$}
    \end{tabular} 
    \endgroup &
    \begingroup
    \renewcommand*{\arraystretch}{0.6}
    \begin{tabular}[c]{@{}c@{}}
        \text{2} \\ 
        {\tiny ICML$^\textit{p}$ NeurIPS$^\textit{p}$}
    \end{tabular} 
    \endgroup &
    \begingroup
    \renewcommand*{\arraystretch}{0.6}
    \begin{tabular}[c]{@{}c@{}}
        \text{2} \\ 
        {\tiny NeurIPS$^\textit{v}$ ICML$^\textit{v}$}
    \end{tabular} 
    \endgroup &
    \begingroup
    \renewcommand*{\arraystretch}{0.6}
    \begin{tabular}[c]{@{}c@{}}
        \text{3} \\ 
        {\tiny EMNLP$^{\textit{v }\mathbf{\dagger}}$ ICML$^\textit{v}$ NeurIPS$^\textit{v}$}
    \end{tabular} 
    \endgroup &
    \begingroup
    \renewcommand*{\arraystretch}{0.6}
    \begin{tabular}[c]{@{}c@{}}
        \text{5} \\ 
        {\tiny FAccT$^{\textit{h }\mathbf{\ddagger}}$ ICLR$^\textit{v}$ ICML$^\textit{h}$} \\ 
        {\tiny NAACL$^\textit{h}$ NeurIPS$^\textit{h}$}\vspace{.5em}
    \end{tabular} 
    \endgroup \\
\textbf{Social Events} & 
    \begingroup
    \renewcommand*{\arraystretch}{0.6}
    \begin{tabular}[c]{@{}c@{}}
        \text{1} \\ 
        {\tiny NeurIPS$^\textit{p}$}
    \end{tabular} 
    \endgroup &
    \begingroup
    \renewcommand*{\arraystretch}{0.6}
    \begin{tabular}[c]{@{}c@{}}
        \text{1} \\ 
        {\tiny NeurIPS$^\textit{p}$}
    \end{tabular} 
    \endgroup &
    \begingroup
    \renewcommand*{\arraystretch}{0.6}
    \begin{tabular}[c]{@{}c@{}}
        \text{5} \\ 
        {\tiny ACL$^\textit{p}$ CVPR$^\textit{p}$ ICML$^\textit{p}$} \\ 
        {\tiny NAACL$^\textit{p}$ NeurIPS$^\textit{p}$}
    \end{tabular} 
    \endgroup &
    \begingroup
    \renewcommand*{\arraystretch}{0.6}
    \begin{tabular}[c]{@{}c@{}}
        \text{11} \\ 
        {\tiny AAAI$^\textit{p}$ AACL$^\textit{v}$ ACL$^\textit{v}$ CogSci$^\textit{v}$} \\ 
        {\tiny  COLING$^\textit{v}$ CORL$^\textit{v}$ EMNLP$^\textit{v}$ FAccT$^\textit{p}$} \\ 
        {\tiny ICLR$^\textit{v}$ \tiny ICML$^\textit{v}$ NeurIPS$^\textit{v}$}
    \end{tabular} 
    \endgroup &
    \begingroup
    \renewcommand*{\arraystretch}{0.6}
    \begin{tabular}[c]{@{}c@{}}
        \text{10}\\ 
        {\tiny AAAI$^\textit{v}$ ACL$^\textit{v}$ CoRL$^\textit{v}$ EACL$^\textit{v}$} \\ 
        {\tiny  EMNLP$^\textit{v}$ ICLR$^\textit{v}$ ICML$^\textit{v}$ NAACL$^\textit{v}$} \\ 
        {\tiny  NeurIPS$^\textit{v}$ \tiny SIGIR$^\textit{v}$}
    \end{tabular} 
    \endgroup &
    \begingroup
    \renewcommand*{\arraystretch}{0.6}
    \begin{tabular}[c]{@{}c@{}}
        \text{7}\\ 
        {\tiny AAAI$^\textit{v}$ AAMAS$^\textit{v}$ ACL$^\textit{h}$ ICLR$^\textit{v}$} \\ 
        {\tiny  ICML$^\textit{h}$ NAACL$^\textit{h}$ NeurIPS$^\textit{h}$}
    \end{tabular}
    \endgroup \\
\bottomrule
\multicolumn{7}{l}{$\mathbf{\dagger}$ at EMNLP 2021, Queer in AI co-hosted a workshop with WiNLP.}\\
\multicolumn{7}{l}{$\mathbf{\ddagger}$ at FAccT, Queer in AI hosted two CRAFT sessions.}
\end{tabular}\\
\smallskip
\small
\noindent
\end{table*}}{\begin{table*}[t]
\caption{Workshop and events organized by Queer in AI in 2017--2022 across conferences in AI. Events marked with \textit{p} were held in person, \textit{v} indicates virtual-only events,  and \textit{h} refers to events that occurred in a ``hybrid'' format. \label{tab:events}}
\footnotesize 
\begin{tabular}{@{}ccccccc@{}}
\toprule
\textbf{Year} & 
    \textbf{2017} & 
    \textbf{2018} & 
    \textbf{2019} & 
    \textbf{2020} & 
    \textbf{2021} & 
    \textbf{2022} \\
\midrule
\textbf{Workshops} & 
    \textbf{-} & 
    \begingroup
    \renewcommand*{\arraystretch}{0.6}
    \begin{tabular}[c]{@{}c@{}}
        \text{1} \\ 
        {\tiny NeurIPS$^\textit{p}$}
    \end{tabular} 
    \endgroup &
    \begingroup
    \renewcommand*{\arraystretch}{0.6}
    \begin{tabular}[c]{@{}c@{}}
        \text{2} \\ 
        {\tiny ICML$^\textit{p}$ NeurIPS$^\textit{p}$}
    \end{tabular} 
    \endgroup &
    \begingroup
    \renewcommand*{\arraystretch}{0.6}
    \begin{tabular}[c]{@{}c@{}}
        \text{2} \\ 
        {\tiny NeurIPS$^\textit{v}$ ICML$^\textit{v}$}
    \end{tabular} 
    \endgroup &
    \begingroup
    \renewcommand*{\arraystretch}{0.6}
    \begin{tabular}[c]{@{}c@{}}
        \text{3} \\ 
        {\tiny EMNLP$^{\textit{v }\mathbf{\dagger}}$ ICML$^\textit{v}$ NeurIPS$^\textit{v}$}
    \end{tabular} 
    \endgroup &
    \begingroup
    \renewcommand*{\arraystretch}{0.6}
    \begin{tabular}[c]{@{}c@{}}
        \text{5} \\ 
        {\tiny FAccT$^{\textit{h }\mathbf{\ddagger}}$ ICLR$^\textit{v}$ ICML$^\textit{h}$} \\ 
        {\tiny NAACL$^\textit{h}$ NeurIPS$^\textit{h}$}\vspace{.5em}
    \end{tabular} 
    \endgroup \\
\textbf{Social Events} & 
    \begingroup
    \renewcommand*{\arraystretch}{0.6}
    \begin{tabular}[c]{@{}c@{}}
        \text{1} \\ 
        {\tiny NeurIPS$^\textit{p}$}
    \end{tabular} 
    \endgroup &
    \begingroup
    \renewcommand*{\arraystretch}{0.6}
    \begin{tabular}[c]{@{}c@{}}
        \text{1} \\ 
        {\tiny NeurIPS$^\textit{p}$}
    \end{tabular} 
    \endgroup &
    \begingroup
    \renewcommand*{\arraystretch}{0.6}
    \begin{tabular}[c]{@{}c@{}}
        \text{5} \\ 
        {\tiny ACL$^\textit{p}$ CVPR$^\textit{p}$ ICML$^\textit{p}$} \\ 
        {\tiny NAACL$^\textit{p}$ NeurIPS$^\textit{p}$}
    \end{tabular} 
    \endgroup &
    \begingroup
    \renewcommand*{\arraystretch}{0.6}
    \begin{tabular}[c]{@{}c@{}}
        \text{11} \\ 
        {\tiny AAAI$^\textit{p}$ AACL$^\textit{v}$ ACL$^\textit{v}$ CogSci$^\textit{v}$} \\ 
        {\tiny  COLING$^\textit{v}$ CORL$^\textit{v}$ EMNLP$^\textit{v}$ FAccT$^\textit{p}$} \\ 
        {\tiny ICLR$^\textit{v}$ \tiny ICML$^\textit{v}$ NeurIPS$^\textit{v}$}
    \end{tabular} 
    \endgroup &
    \begingroup
    \renewcommand*{\arraystretch}{0.6}
    \begin{tabular}[c]{@{}c@{}}
        \text{10}\\ 
        {\tiny AAAI$^\textit{v}$ ACL$^\textit{v}$ CoRL$^\textit{v}$ EACL$^\textit{v}$} \\ 
        {\tiny  EMNLP$^\textit{v}$ ICLR$^\textit{v}$ ICML$^\textit{v}$ NAACL$^\textit{v}$} \\ 
        {\tiny  NeurIPS$^\textit{v}$ \tiny SIGIR$^\textit{v}$}
    \end{tabular} 
    \endgroup &
    \begingroup
    \renewcommand*{\arraystretch}{0.6}
    \begin{tabular}[c]{@{}c@{}}
        \text{7}\\ 
        {\tiny AAAI$^\textit{v}$ AAMAS$^\textit{v}$ ACL$^\textit{h}$ ICLR$^\textit{v}$} \\ 
        {\tiny  ICML$^\textit{h}$ NAACL$^\textit{h}$ NeurIPS$^\textit{h}$}
    \end{tabular}
    \endgroup \\
\bottomrule
\end{tabular}
\smallskip
\small
\noindent
\begin{itemize}
    \item[$\mathbf{\dagger}$] at EMNLP 2021, Queer in AI co-hosted a workshop with WiNLP.
    \item[$\mathbf{\ddagger}$] at FAccT, Queer in AI hosted two CRAFT sessions.
\end{itemize}
\end{table*}}

\subsubsection{Workshop Organizing} 
Queer in AI workshops and socials are typically organized by members of the community planning to attend the conference; no prior academic or organizing experience is required. Junior or new members of the community are often encouraged to lead these initiatives while being mentored by more experienced organizers throughout the process. Organizers, DEIAs, and Queer in AI’s financial stewards coordinate to secure logistical, monetary and other miscellaneous needs of the event. These include renting equipment to support accessibility, honoraria for speakers, scholarships for attendees, refreshments for socials, online outreach and promotion of the event, and so on. All of this communication takes place asynchronously over Slack, or in Zoom meetings scheduled across organizers' time zones. This decentralized approach also helps enable Queer in AI members spanning different sub-fields in AI to tailor events to represent and serve the needs of their sub-community. When prompted to rate how welcome they felt at these workshops, the response was overwhelmingly positive, with about 47\% of queer attendees rating it five out of five on a Likert scale ($\mu$=4.16, $\mu_{1/2}$=4) (\S\ref{ssec:demosurvey}).

\subsubsection{Panels and Talks at Workshops}
Panels and talks at Queer in AI are crucial as they help in amplifying queer voices and concerns in our field. 
Many topics presented in the panels and keynotes have later served a bigger impact in the AI field, such as talks on conference inclusivity and name change policies.
Queer in AI encourages a participatory approach to workshop design: by soliciting topics and speaker ideas from community workspace.
This approach has allowed Queer in AI to host panels and talks on intersectional topics that often do not have a presence at major AI/ML venues (for just one example, a discussion on the intersection of queerness, caste and AI at NeurIPS 2021~\cite{qaineurips2021}).
Queer in AI organizers spend tremendous effort by making the workshops as inclusive as possible by providing fair honoraria to the speakers and organizing the events in online, hybrid, and in-person settings. 

\subsubsection{Barriers and Challenges in Participation}
\label{subsub:conf_participation_challenges}
AI conferences are often not accessible for a sizable portion of queer researchers, especially those belonging to other marginalized backgrounds or from countries with lower purchasing power or higher rates of discrimination towards queer people~\cite{tulloch2020improving}.
Primary reasons includes high registration and travel costs.
Out of all Queer in AI members who reported being unable to attend conferences owing to lack of funding, 88\% identified as one of Black, indigenous, person of color, transgender, neurodivergent, or disabled (\S\ref{ssec:demosurvey}).
While Queer in AI tries to work with conference organizers to use DEI funds for increasing the attendance of queer scientists, in many cases conference organizers refuse to engage with Queer in AI’s requests. Queer in AI thus often provides a combination of travel grants, registration waivers, and reimbursement for conference-related expenses to queer AI researchers. In other cases, unofficial social events\footnote{These events are not officially included within the conference program but promoted over Queer in AI's Slack and mailing list as well as social media. A recent example is AAAI 2023 where the conference fees was exorbitantly high and negligible effort was put into provision for registration waivers.} near the conference venue and online virtual socials on \href{https://gather.town}{gather.town} are organized to accommodate excluded time zones and overcome both financial and geographical access barriers.
Other barriers specific to the conference location, such as unsafe legal and social climates~\footnote{EMNLP 2022 (in Abu Dhabi) predatorily included Queer in AI to obtain their approval for conference safety measures; Queer in AI rejected this, due to the conference operating at a different domain of power for trans people and the power inherent in speaking for the entire queer community.} for queer people or exclusionary visa processes, continue to significantly limit queer participation within AI spaces.
Finally, for conferences which are poorly equipped in their support for disabled people, Queer in AI provides live captions for all in-person and virtual events, and secures equipment to create accessible spaces.

\subsection{Advocacy for Improving Queer Inclusivity in Conferences}
\label{sub:inclusive_conf_guide}

As conferences moved online in response to the COVID-19 pandemic, Queer in AI organizers noted a series of operational failures that could cause queer attendees to feel unsafe or unwelcome. 
Registration platforms demanded attendees to provide their legal names, thus potentially deadnaming them; 
the use of pronoun badges for speakers and attendees was rarely encouraged, or platforms did not support displaying pronouns;
virtual chat software blocked common queer terms such as ``queer'' or ``lesbian'', thus preventing queer attendees from communicating freely.
Queer in AI organizers worked closely with many conferences to resolve these issues,
as they had in prior settings (\S\ref{sub:workshops}),
and ultimately decided to collect recommendations
aimed at highlighting best practices to ensure safety, privacy, and accessibility for queer attendees at academic conferences in AI
in a collected guidance document.\footnote{The guide, originally published as \cite{queer-in-ai-dni-guide}, is a living document available at \href{http://queerinai.com/how-to-make-virtual-conferences-queer-friendly}{queerinai.com/how-to-make-virtual-conferences-queer-friendly}.}

These recommendations began based on existing best practices and experience with conference organizers, but were refined through extensive iterative feedback from members of Queer in AI and other affinity groups, incorporating many opinions and ultimately achieving consensus among a broad group of contributors.
The guide has recently been expanded to also cover in-person events as conferences move to hybrid or in-person formats. 
This queer advocacy to improve inclusivity covers two aspects: improving queer safety and increasing queer representation.

\subsubsection{Improving Queer Safety:}
As in any public space, queer conference-goers might face discrimination based on their gender and sexual orientation. 
Therefore, it is paramount for attendees to be able to control what information they wish to disclose to the organizers and attendees of a conference. 
Queer in AI advocates mechanisms to (i) respect attendees' identities by collecting gender and pronoun information in a manner that does not misrepresent or erase queer identities, by creating forms with inclusive gender categories and disclosing the data usage~\cite{morgan_gender} (ii) minimize the amount of personal information queer individuals have to disclose~\cite{data-collection} (for example, only collecting legal names when absolutely necessary, and using responses about the gender and sexuality of attendees only for statistical purposes and in anonymized form); and (iii) ensure that mechanisms to report disruptive or harmful behaviours are swift and effective. Queer in AI recommends adopting a code of conduct (\textit{e.g.}, \cite{queerinai_code_of_conduct,wiml_code_of_conduct}) to not only establish communication norms, but also describe how policy violations are handled~\cite{ashedryden2014}.

\subsubsection{Increasing Queer Representation and Participation:}
Queer 
researchers' needs are regularly ignored in many aspects of the research community: challenges include lack of academic support, hostility from colleagues and advisors, inflexible name change policies, lack of representation in the research itself, and more~\cite{socsci6010012}.
Stronger inclusion efforts, both for representation and participation, can work towards addressing a lack of queer community and role models~\cite{schluter2018glass}.
To increase representation, Queer in AI strongly encourages conference organizers to invite queer keynote speakers and panelists, prioritizing those from marginalized backgrounds (\textit{e.g.}, BIPOC or non-cisgender) \cite{ashedryden2013}.
Queer in AI recommends fair and equal compensation based on effort rather than seniority for all speakers~\cite{reid2021speaker, forbes2021guadiano}.
As noted in previous sections, financial accessibility and a lack of community were the main barriers for queer folks to feel included at conferences.
Queer in AI strongly advocates setting up spaces for queer folks to network and socialize with privacy measures and also providing subsidies for queer researchers to attend virtual or in-person events.

\subsubsection{Critical Reflection}
This guide and advocacy are not without their limitations.
Most recommendations are still focused on virtual spaces and currently written guide lacks in-depth accessibility recommendations.
Queer in AI needs to collaborate with disabled folks with a wider range of disabilities to document best practices regarding accessibility accommodations.
Most significantly, despite organizers' efforts the guide has seen relatively modest adoption. 

\subsection{Trans-inclusive Publishing Advocacy}
\label{sub:namechange}

For many transgender, non-binary, and gender-diverse scholars (as well as others), the continued circulation of a previous name in publishing is a significant source of trauma~\cite{cope-vision}. 
Referring to an author by a previous name without consent (deadnaming) may effectively out their identity against their will. 
Queer in AI has worked along with the Name Change Policy Working Group \cite{ncpwg} to advocate name change policies in AI venues,
helping to establish the name-change policies and procedures now adopted by most AI-related venues~\cite{aclanthocorrections,naaclnamechange,ieeenamechange,acmnamechange,neuripsnamechange,pmlrnamechange,openreviewnamechange, arxivnamechange} (cf. \S\ref{sec:policy} for more about Queer in AI's advocacy and impact).

Even publishers with functional name change policies
are often woefully slow to implement them,
and search engines can index outdated information long after its correction \cite{robyn_doc,scholarfailed};
moreover, authors often use outdated bibliographic entries long after relevant publications and search tools have been updated \cite{danicaslides}.
It is thus vital to check the correctness of citations in submitted papers to avoid propagating incorrect information.
QueerInAI has thus developed a tool to check paper PDFs for mistaken citations.
It searches the ACL Anthology, DBLP, and arXiv for a close paper title match,
and prompts a correction if the paper's author list disagrees with that source,
detecting both deadnaming and incomplete or outdated author lists.
DBLP in particular provides better name change support than many other platforms, via ORCID \cite{orcid}.
This toolkit has been integrated into ACL publication camera-ready systems~\cite{aclpubcheck}, and Queer in AI hopes to expand it to other conferences.
A demo is available at \href{http://qinai-name-check.streamlit.app}{qinai-name-check.streamlit.app}.

Additionally, Queer in AI advocates publishers to promptly grant name correction requests in any format, without unnecessary barriers or documentation requirements.
Such changes should remove all instances of authors' previous names from all records, or (at the author's discretion) add disclaimers for media that cannot be updated (\textit{e.g.}, audio or video recordings). 
As the result of this advocacy, Queer in AI has helped institute effective name-change processes at NAACL and EMNLP;
and has worked with the Association for Computational Linguistics \cite{aclweb} to implement a name change process, proactive measures to prevent the deadnaming of trans authors, and protocols to handle authors' requests to keep their videos private.

\section{Tensions and Challenges} \label{sec:discussion}
As reflexivity is a core tenet of intersectionality \cite{collins2019intersectionality}, this section critically examines the tensions and challenges that emerge in the operationalization of Queer in AI's principles within its initiatives. From the issues with Queer in AI initiaitves discussed in the previous section, we find three common, root themes of \textbf{hierarchy, accessibility}, and \textbf{funding}. We argue that these are not only critical challenges for Queer in AI, but deep challengesany participatory or community-lead AI organization must address to be successful. 

\subsection{Hierarchy}
Decentralized organizing plays a vital role in minimizing power distance and distinctions between members of Queer in AI. 
Even so, there are notable distinctions between members who participate in organizing, core organizers, and the DEIAs as paid contractors. 
Queer in AI’s core organizers and DEIAs help sustain the growth of the organization through mentorship of new volunteers and institutional memory. In addition, they form a relatively large and diverse group for deliberating on rare decisons that cannot be discussed openly, such as those involving PII. Their existence does, however, pose challenges in accessibility for people unfamiliar with navigating unstructured social networks, and can be non-transparent to newer or less involved members. 
The core organizers also assume a more active role, sharing considerable power in steering the direction of its initiatives. 
Queer in AI helps address these tensions by setting a fixed one-year tenure for DEIAs, and inducting organizers who have been active throughout the preceding year as core organizers. Resolving tensions between decentralization and hierarchies created by knowledge and experience, or forced by privacy concerns, nonetheless remains an open problem within Queer in AI.

\subsection{Accessibility}  \label{subsub:access_to_organizing}
Despite global participation, Queer in AI's structure and operational design can discourage participation for many queer scientists. 
First, participation in a volunteer-run community not only requires organizers to have income that allows them to perform free labor but also have access to computers, internet, and other resources required to even connect with Queer in AI. 
Second, while Queer in AI strives to be intersectional, it severely lacks access to queer networks in countries from the Global South. It originated and primarily operated within a Western context during its initial years, which led to the inadvertent creation of barriers that limit its outreach.
For example, because Queer in AI organizers are best connected with US and European institutions, its events are often co-located at conferences attended mainly by scientists residing in the Global North. 
Further, its meetings often occur at times best aligned with European and American time zones, at the expense of much of Asia. Finally, all Queer in AI activities require English proficiency.

While recent community and focused outreach efforts have reduced some of these barriers, significant work lies ahead in establishing truly global ways of participation, especially for countries where queerness is criminalized. Third, participation in Queer in AI exerts a toll on mental health and exhaustion of its organizers (Figure~\ref{fig:organizer_challenges}). This is partly due to Queer in AI's lack of formal structure, instead relying on individuals self-coordinating on initiatives of their choice. While efficient, this approach can make joining and keeping track of ongoing efforts challenging for newcomers and neurodivergent members of the community. Past organizers have also shared anecdotes of experiencing exhaustion, fatigue, and anxiety due to a lack of accommodation of different working styles and falling behind on personal schedules while undertaking operational work for Queer in AI (see Figure~\ref{fig:organizer_challenges}). This disproportionately impacts disabled and neurodivergent members and is compounded for intersecting marginalized identities.

Even after years of critical reflection and significant investment of volunteer time, money, and other resources, Queer in AI is still inaccessible to many. 
While accessibility to everyone should always be the goal, in practice, no single community or participatory initiative will be able to include everyone in that community. Therefore, participatory researchers aspiring to broad inclusion should consider the pluralities of communities and participatory initiatives with radically different structures.

\subsection{Funding}  \label{sec:funding}
Funding and payments are where Queer in AI struggles most to meet its commitments to decentralization, intersectionality, and community leadership. 
Queer in AI relies on sponsorships, donations, and contributions from its parent organization oSTEM to fund
its activities.
In 2022, Queer in AI expenses (rounded to the closest integer) totaled US\$100,658:
the graduate application fee scholarship program (\S\ref{sec:grad-app}) spent \$40,435;
two DEIA contractors were paid a total of \$33,220;
speaker honoraria totaled \$14,500;
\$6,941 went to travel grants, room and board, and conference registration fees;
emergency microgrants for queer people totaled \$5,000.
Income comprised \$78,000 in corporate sponsorship, \$13,711 in donations, and \$5,000 in grant revenue
(cf.\ Appendix~\ref{app:finances} provides income and expenses for previous years.).

Queer in AI's reliance on corporate sponsorship may call into question its independence and community-lead ideal.
Corporate sponsors receive access to opt-in resume books, short speaking opportunities, and event recruiting booths.
A large part of Queer in AI's funding still comes from big tech corporations that are complicit in oppression and genocide globally, such as the policing of Palestinians.
Queer in AI has nonetheless dropped and turned down many sponsors for ethics concerns, including a mutual decision with Black in AI in 2021 to drop Google~\cite{khari2021drop}, costing \$20,000 in lost sponsorship per year.
While Queer in AI has been growing donations, many in the Queer in AI community are students or early in their careers with very limited capacity to give. Opportunities for grants are limited, as many scientific funding bodies such as the US's NSF exclude queer people from many of their D\&I initiatives~\cite{freeman2023}.

Queer in AI sends honoraria, scholarships, and travel grants to people in many different countries, primarily through PayPal and wires. Payment disbursal in Queer in AI is highly centralized; for reasons of security oSTEM only allows one Queer in AI organizer to send PayPal payments. All wires and credit card payments must be sent by the oSTEM CEO. Additionally, payments strain Queer in AI's intersectional values. PayPal does not work well in China, India, many countries in Africa, and some countries in South America, forcing reliance on slower and more administratively difficult wire transfers.
Moreover, U.S. law requires people receiving honoraria and other types of payments to pay US taxes above a certain threshold, which requires a lengthy registration process or significant fees and overhead from Queer in AI.
Payments also frequently trigger spurious fraud alerts and investigations, which require even more time from and stress on organizers.

In summary, marginalization prefigures Queer in AI's funding options, legal and security concerns exert a strong centralizing pressure on financial administration, and the financial system regards many payments, especially to non-Western countries and those making them, with suspicion by default.

\section{Conclusion} \label{sec:conclusion}
Participatory methods have the potential to address issues of power and inclusion in AI, but their benefits and challenges in practice are still unclear because few organizations have deeply engaged with them. In this paper we studied Queer in AI as a case study of a grassroots participatory AI organization. We explored how they designed their organization to enable participation, and how initiatives addressing intersectional marginalization arose from and were continuously refined by this participation. We theorized how Queer in AI's numerous socials, workshops, and other events have contributed to a culture of participation in AI by bringing queer people into AI conferences and research and industry settings and resisting predatory inclusion.
We hope this case study will inform theoretical study and practical design of participatory initiatives. In particular, we encourage consideration of Queer in AI's reinforcing principles of decentralization, community leadership, and focus on intersectionality, and urge care for mitigating the ways hierarchy, inaccessability, and funding can subvert participatory methods.

\subsection{Future Directions}
Queer in AI will continue to grapple with the tensions and alleviate the challenges addressed in \S\ref{sec:discussion}. To dismantle hierarchies among organizers created by knowledge, experienced Queer in AI organizers will host structured trainings to onboard new organizers onto finance \& sponsorships and workshops. Queer in AI additionally plans to supplement its 2023 community survey with community interviews about accessibility, towards gleaning actionable insights about mitigating barriers to participation. Furthermore, Queer in AI's organizers will work with its community to refine its sponsorship policies and identify less precarious mechanisms for transferring funds. All of these activities are motivated and will be guided by our core principles of decentralization, intersectionality, and centering community. Queer in AI will further communicate its activities and their implications for equity and inclusivity via accessible media, e.g., blog posts, zines.

\begin{acks}
This work would not have been possible without the activism and organizing efforts of the Queer in AI community. We would also like to thank Katta Spiel and Os Keyes for their insightful feedback on the earlier versions of the paper. 
\end{acks}

\bibliographystyle{ACM-Reference-Format}
\bibliography{sample-base}

\clearpage
\appendix
\renewcommand{\thetable}{A\arabic{table}}
\renewcommand{\thefigure}{A\arabic{figure}}
\setcounter{figure}{0}
\setcounter{table}{0}

\section{The Limits of Computational Methods to Achieve Fairness}
\label{app:fairness_limitations}

Computational approaches to fair machine learning ensure that models satisfy certain mathematical formulations of fairness or don't capture social biases. For example, a fair model may treat similar individuals similarly (individual fairness), predict similar outcomes across sensitive attributes (group fairness), or learn representations that are high-quality for all individuals and don't encode stereotypes related to sensitive attributes ~\cite{corbett2018measure, 10.1145/3461702.3462621}. Different operationalizations of fairness in machine learning encode different theoretical understandings of fairness, and many operationalizations, due to their quantitative nature and focus on parity, don't capture notions of fairness based in, for example, representational justice \cite{jacobs2021measurement}. Many researchers consider vectors of unfairness in machine learning models to include tainted examples (e.g., historical data that capture stereotypes and discrimination), limited features (e.g., unobserved features for marginalized communities), and sample size disparities (e.g., significantly fewer data for minoritized groups) \cite{barocas-hardt-narayanan}. However, researchers often neglect how these vectors result from interlocking power relations and the social inequality that these relations produce \cite{kong2022intersectional}.

Current paradigms for ensuring fairness in machine learning largely rely on historical data and observed attributes \cite{barocas-hardt-narayanan}. Causality is emerging as a lens through which fairness can be observed under intervention \cite{barocas-hardt-narayanan, NIPS2017_a486cd07}, but it too assumes that sensitive attributes and identities are \textbf{known}, which is often not the case due to privacy laws and the dangers involved in disclosing certain sensitive attributes (e.g., disability, queerness, etc.) \cite{theilen2021protection, gdprpersonaldata}; \textbf{measurable}, which is almost never true (e.g., gender) \cite{dev2021harms}; \textbf{discrete}, which reinforces hegemonic, colonial categorizations (e.g., race and ethnicity options on the US/UK census, the gender binary, etc.) \cite{hanna2020race, onscensus}; and \textbf{static}, which is problematic given that one’s identity can change over time (e.g., genderfluidity) \cite{dev2021harms}. These assumptions especially pose problems in the context of queerness since sexuality and gender identity are often ``unobserved characteristics, which are frequently missing, unknown, or fundamentally unmeasurable'' \cite{tomasev2021fairness}. Furthermore, observational fairness neglects that some communities face complex, intersecting vectors of marginality that preclude their presence in the very data observed for fairness.

Another pitfall of fair machine learning is the tendency to mitigate biases in models posthoc; any model that is created without queer people in
mind and is band-aided after the fact to protect them from harm will still inevitably harm them. Additionally, to improve model fairness for queer communities, current paradigms of fairness dictate to collect more data on them, which exposes queer people to predatory inclusion, serious privacy risks, and even violence \cite{agnew2021rebuilding, Cooper_2021, seamster2017predatory, beyondfairness2021, geeng2021lgbtq}. Data are often not collected consensually \cite{PAULLADA2021100336}. Furthermore, given the dire underrepresentation of queer people in datasets to begin with, without caution, even simple demographics can uniquely identify queer persons \cite{sweeney2000privacy}. This could be disastrous for queer people living under violent, oppressive institutions.

Not all fair machine learning models distribute justice to queer communities. For instance, applications of fair link prediction in social networks to deliver content or connection recommendations that are independent of users' identities could be problematic \cite{li2021on, subramonian2022ondyadicfairness}. Many LGBTQIA+ people create and rely on the sanctity of safe spaces online \cite{harper2016internet}. Thus, recommending them users or news sources that are hostile (e.g., promote homophobic, racist, or sexist content) can result in severe psychological harm and a violation of privacy. Furthermore, many queer individuals feel isolated in real life and actually yearn to find other users online who share their identity, to which fair link prediction is antithetical \cite{aven}. Moreover, fairness does not benefit a model that is inherently flawed. Researchers have attempted to build neural networks to infer sexuality from images of people, however, sexuality is fundamentally not detectable by a human or machine learning model~\cite{newclothes}. As such, no amount of fairness can compensate for the reality that the premise of the model is based in physiognomy \cite{stark2021physiognomic} and assumes the biological essentialism of sexuality and expression. Furthermore, a sexuality detection system, regardless of how effective, may be easily weaponized by oppressive institutions against queer and cishet people alike, with severe representational and allocational harms, from violation of privacy to death. Fairness cannot help such a system intended to police queer bodies.

Finally, time after time, institutions that claim to act without discriminating on the basis of protected class can produce disparate impact \cite{feldman2015disparate}. Similarly, machine learning models that purportedly make decisions independently of sensitive attributes automate systemic oppression. A model that assists and benefits queer people will take into account queer identities and actively strive to improve societal equity.

\section{Queer in AI Surveys}
\label{app:surveys}

This section overviews the creation of and methodologies Queer in AI employs in administering its demographic survey. Following this, survey results for Queer in AI's organizers and community are presented.

\subsection{Queer-Inclusive Data Science}

Data science and analytics have historically been weaponized against marginalized communities, including queer people, by justifying policies that permit or amplify inequality and discrimination \cite{redden2020dataharm}. Data collection often also poses serious privacy and security risks to queer communities \cite{redden2020dataharm, gutierrez2018privacy}.
Queer in AI attempts to, in part, reclaim data science as a tool for justice. Queer in AI administers an organizer survey (\S\ref{ssec:organizersurvey}) to understand organizers' identities, their motivations for and obstacles to volunteering, and other issues that they face, as well as a related community survey (\S\ref{ssec:demosurvey}).

\subsection{Curation Rationale}

The responses to Queer in AI's organizer survey\footnote{\url{https://forms.gle/oSHTtpkdzUvpNhQL6}} are used to better understand organizers’ identities, their motivations for and obstacles to volunteering, and other issues that they face. The community survey\footnote{\url{https://docs.google.com/forms/d/e/1FAIpQLSes-lzwkKHruQrAmH3Tnz1tJsTUl-YP51V8wDtHbfb8Z9FoNg/viewform}} is used to identify issues within queer communities, shape the future programs of Queer in AI, and inform its operation. In particular, Queer in AI uses the community survey for: 
\begin{itemize}
\item understanding issues and status of queer communities,
\item understanding queer intersectionality
\item collecting info on trans-inclusive publications,
\item collecting info on queer inclusivity in academia and conferences,
\item shaping Queer in AI mentoring programs,
\item getting feedback on Queer in AI socials and initiatives. 
\end{itemize}

\subsection{Data Collection Policy}

The organizer survey is only sent out to Queer in AI organizers via Slack. The community survey is sent out to Queer in AI members via Slack, all attendees of Queer in AI socials and workshops, and is placed on top of Queer in AI's website. Respondents can be anonymous while entering the survey responses. All questions in both surveys are optional; in demographic questions like gender and sexual orientation, respondents can choose multiple responses; and many questions allow free-text responses. Only a handful of Queer in AI organizers have access to these responses. Folks can contact Queer in AI to delete the information if they want. LGBTQ Crisis Hotlines are linked in the survey considering the nature of the questions.

\subsection{Survey Curators' Demographics} The surveys were designed in collaboration with gender theory scholars and transgender, gender-diverse and BIPOC members of Queer in AI. All of the curators have informal training in queer studies through activism and advocacy in Queer in AI and affiliated groups. The curation team consisted of 8 members. They ranged in age from 21-35 years, with gender including men (2), women (2), non-binary folks (2), agender individual (1) and genderfluid (1). 5 of them are transgender. 5 of them are BIPOC. Region-wise, 1 is from East Asia, 1 is from Europe, 1 is from South Africa, 1 is from South America and 4 are from North America.

\subsection{Queer in AI Organizers} \label{ssec:organizersurvey}

Decentralized, participatory organizations are in large part defined by the people who volunteer to run events and initiatives.
To empower and represent the queer community,
in all its diversity (not just across gender and sexual and romantic orientations, but also disability, ethnicity, place of origin, economic background, and many more)
it is vital to
recruit, train, and retain a diverse set of volunteers. This section discusses the demographics of Queer in AI's volunteers and Queer in AI's experiences in recruiting and training their volunteers.

\begin{figure*}[htbp]
\centering
\subfloat[Sexual orientation of Queer in AI organizers.] {
    \includegraphics{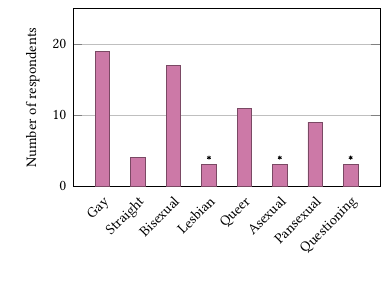}
    \label{fig:org_so}
}\hspace{1.5cm}
\subfloat[Gender of Queer in AI organizers.] {
    \includegraphics{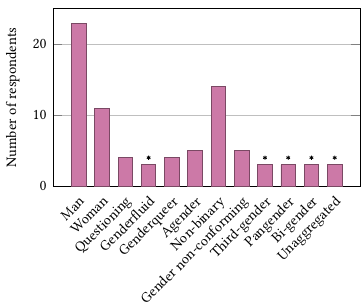}
    \label{fig:org_gender}
  }
\caption{Sexual orientation (\subref{fig:org_so}) and gender (\subref{fig:org_gender}) statistics of the Queer in AI organizers (2021--2022). Write-in responses were aggregated by a team of Queer in AI organizers, with some falling into multiple categories. ``Unaggregated’' refers to responses that could not be adequately described with any subset of other categories; however, responses in this group may overlap with the remaining categories. For categories with $\leq3$ responses (marked by *), exact numbers are omitted to protect anonymity. \label{fig:org_gender_and_so}}
\end{figure*}

\begin{figure*}[t]
\centering
\includegraphics{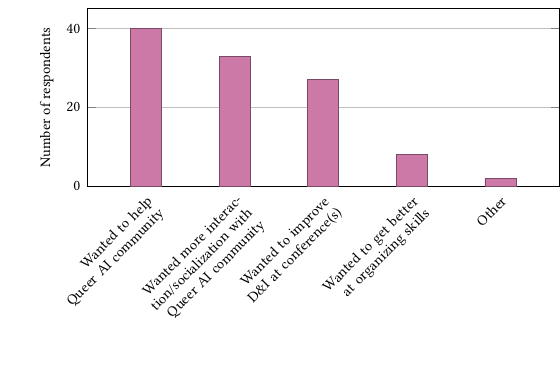}
\caption{Motivations that Queer in AI organizers had for volunteering. \label{fig:organizer_motivation}}
\end{figure*}

\begin{figure*}[t]
\centering
\includegraphics{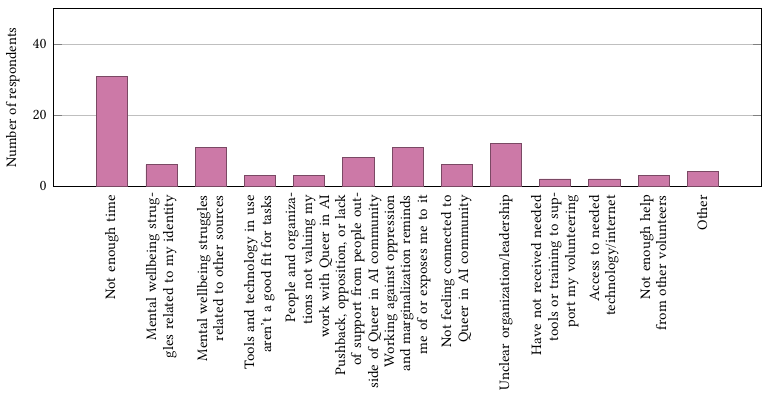} 
\caption{Challenges that Queer in AI organizers faced when volunteering. \label{fig:organizer_challenges}}
\end{figure*}

\begin{figure*}[t]
\centering
\includegraphics{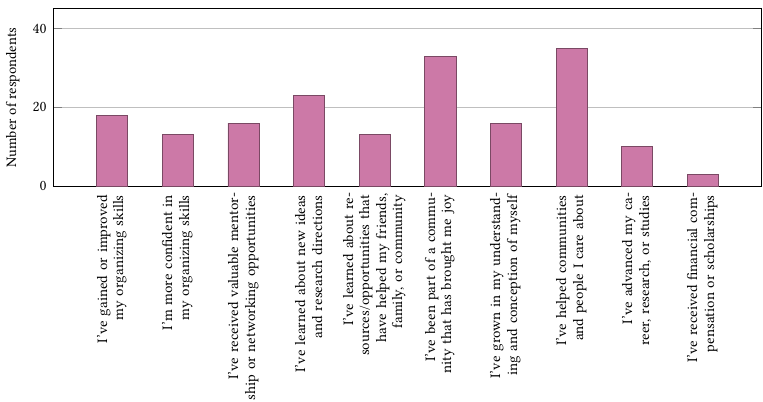}
\caption{Benefits that Queer in AI organizers received from volunteering. \label{fig:organizer_benefits}}
\end{figure*}

\begin{figure*}[t]
\centering
\includegraphics{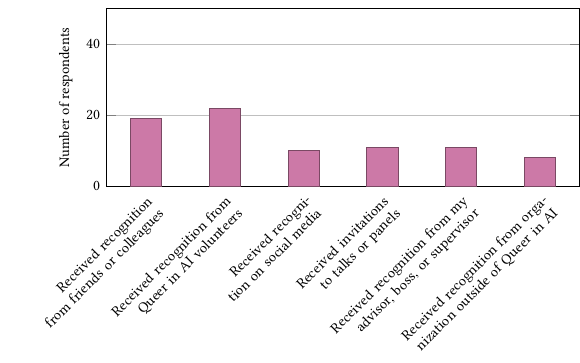}
\caption{Recognition received by Queer in AI organizers. \label{fig:organizer_recognition}}
\end{figure*}

Figures \ref{fig:org_so}, \ref{fig:org_gender}, \ref{fig:org_co}, \ref{fig:org_cl}, and Table \ref{tab:org_career} present demographics of Queer in AI's organizers based on a recent survey. To protect anonymity, categories with less than or equal to three responses are marked by *. Most organizers identify as gay, bisexual, and/or queer, and as men and/or non-binary; lesbians and women remain underrepresented. Almost all volunteers wanted to help the community, but 75\% also wanted community, highlighting the importance of socialization and fun in organizing (Figure ~\ref{fig:organizer_motivation}). By far the biggest challenge Queer in AI's volunteers faced (Figure \ref{fig:organizer_challenges}) was not having enough time (70\%).
While most volunteers have received recognition of their work from friends, colleagues, or other volunteers, only 25\% had received recognition from their bosses or advisors, and none had received awards or other such recognition~(Figure \ref{fig:organizer_recognition}).
The time pressures and lack of outside recognition reflects a lack of institutional support for Queer in AI: its volunteers \textit{want} to contribute more, but their jobs and careers will not support or value them doing so.
Even so, volunteering with Queer in AI has helped many of its organizers professionally, with most reporting getting the experience to organize socials, workshops or conferences, and many reporting finding research opportunities or help applying for jobs or school through Queer in AI, both valuable connections especially given that a majority of its organizers are students or early-career.
85\% of Queer in AI's organizers report that volunteering has allowed them to help the communities they care about and 75\% report being a part of a community that has brought them joy, showing that Queer in AI is successfully providing volunteers with the things they joined for by providing a space to build solidarity, understanding and learning from each others experience/issues (Figure \ref{fig:organizer_benefits}).

\begin{table}[h]
\begin{tabular}{lr}
\toprule
    Undergraduate student & $\leq$3 \\
    Graduate student & 15 \\
    Junior academic & 6 \\
    Senior academic & 8 \\
    Junior industry & $\leq$3 \\
    Senior industry & 6 \\
    Other & $\leq$3\\
\bottomrule
\end{tabular}
\caption{Queer in AI organizers' career stages}
\label{tab:org_career}
\end{table}

However, Queer in AI's organizers lack diversity along several important axes, including place of origin, ethnicity, gender identity (including trans identities), caste, neurodivergence and disability.
Recent programming at workshops has featured a variety of talks and panels centering trans, non-binary, and queer BIPOC issues, including discussions of how Queer in AI can do better. Efforts are also continuing to help encourage community members to raise any concerns about inclusivity, including feedback on the community survey discussed next and through other channels.

\begin{figure*}[htbp]
\centering
\includegraphics{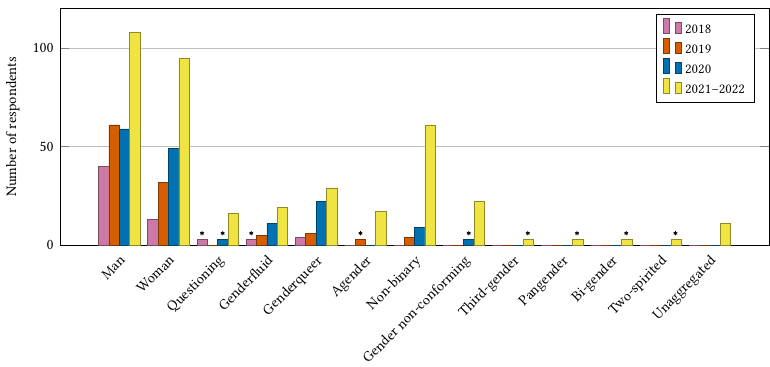}
\caption{Gender statistics of the Queer in AI community members. Data was collected via the demographic surveys (\S\ref{ssec:demosurvey}). Write-in responses were aggregated by a team of Queer in AI organizers, with some falling into multiple categories. ``Unaggregated’' refers to responses that could not be adequately described with any subset of other categories; however, responses in this group may overlap with the remaining categories. For categories with $\leq3$ responses (marked by *), exact numbers are omitted to protect anonymity. \label{fig:comm_gender}}
\end{figure*}

\begin{figure*}[hbtp]
\centering
\includegraphics{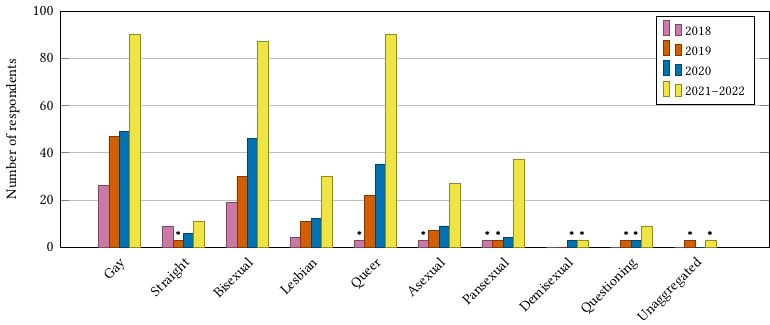}
\caption{Sexual orientation statistics of the Queer in AI community members. Data was collected via the demographic surveys (\S\ref{ssec:demosurvey}). Write-in responses were aggregated by a team of Queer in AI organizers, with some falling into multiple categories. ``Unaggregated’' refers to responses that could not be adequately described with any subset of other categories; however, responses in this group may overlap with the remaining categories. For categories with $\leq3$ responses (marked by *), exact numbers are omitted to protect anonymity. \label{fig:comm_so}}
\end{figure*}

\subsection{Queer in AI Community} \label{ssec:demosurvey}

This section presents five years of demographics for the Queer in AI community.
To protect the anonymity of survey participants, exact numbers are reported only for categories with more than three responses (categories with $\leq3$ are marked by *). Overall, the community has grown significantly since 2018, with the number of survey respondents growing by almost five-fold.

\begin{figure*}[hbtp]
\centering
\subfloat[Queer in AI Organizers (2021--2022): Self-Reported Ethnicity.] {
    \includegraphics{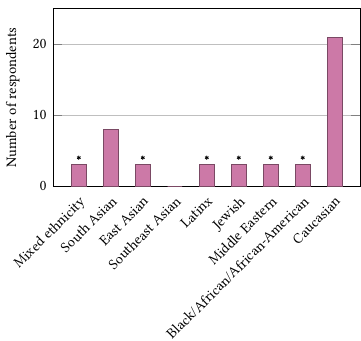}
    \label{fig:org_eo}
}\hspace{2cm}
\subfloat[Queer in AI Scholarship Recipients: Self-Reported Ethnicity.] {
    \includegraphics{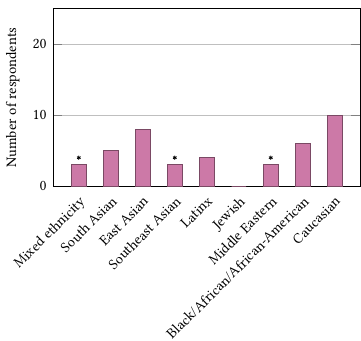}
    \label{fig:org_esr}
  }\\
\subfloat[Queer in AI Community (2021--2022): Self-Reported Ethnicity.] {
    \includegraphics{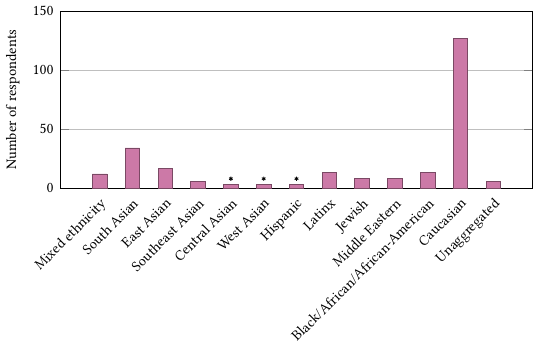}
    \label{fig:org_ec}
}
\caption{Self-reported ethnicities of Queer in AI's organizers (\subref{fig:org_eo}), scholarship recipients (\subref{fig:org_esr}), and community members (\subref{fig:org_ec}) respectively. Write-in responses were aggregated by a team of Queer in AI organizers, with some falling into multiple categories. ``Unaggregated'' refers to responses that could not be adequately described with any subset of other categories; however, responses in this group may overlap with the remaining categories. For categories with $\leq3$ responses (marked by *), exact numbers are omitted to protect anonymity.}
\end{figure*}

\paragraph{Demographics} In terms of sexual orientation, most of Queer in AI members identify as gay, bisexual, or queer, with the latter category increasing significantly in recent years; lesbians remain underrepresented in the Queer in AI community (Figure~\ref{fig:comm_so}). 
While 2021 saw a significant increase in non-binary members, men and women remain the most common gender identities in the community (Figure~\ref{fig:comm_gender}). 
Since 2020, Queer in AI has also surveyed its members about their ethnicity (Figure~\ref{fig:org_ec}); community members overwhelmingly self-describe as white, followed by South Asian, Latinx, and East Asian. 
About 22\% of survey respondents identify as transgender,
10\% as disabled, and 30\% as neurodivergent.

\paragraph{Safety} Despite recent worldwide progress in queer rights, Queer in AI members still face significant discrimination (Table~\ref{table:safety}). 67\% of members reported to have faced at least one safety incident in 2021, the most common being target of jokes and innuendos (47.9\%), being deliberately ignored or excluded (43.6\%), or being singled out as resident authority (42.1\%). Members who live in countries where queer people are persecuted have a similar distribution of incidents, but at a much higher rate. BIPOC members reported higher rates of incidents, with 38\% mentioning facing microaggressions.  

\paragraph{Mental Health} An overwhelming majority of survey respondents reported mental health hardships (Table~\ref{table:mental_health}). Many reported that mental health issues have impaired their ability to conduct research (79.9\%), especially neurodivergent members (91.9\%); About a third reported that they have harmed themselves or considered suicide. 
Results in Table~\ref{table:mental_health_2} show that Queer in AI members struggle the most with their mental health as a student (43.2\%); over a third has reported struggles in the last year, perhaps related to the continued impact of the COVID-19 pandemic. 

\paragraph{Obstacles}
Members of the Queer in AI community also struggle with a lack of community they can rely on (77.4\%). Undergraduate students are especially affected, with 91.6\% reporting a lack of a support group they could rely on. 
On a 1--5 scale, members reported a lack of representation of non-cisgender (1.7~/~5, Table~\ref{table:noncis_fam}), and BIPOC folks (2~/~5) in their immediate work environment (Table~\ref{table:bipoc_fam}).
Cisgender survey respondents remain largely unaware of specific issues affecting non-cisgender scholars (Table~\ref{table:cis_fam}), such as lack of name changes policies for many academic journals.
Finally, slightly more Queer in AI members recently reported having come completely out: 47\% in 2021, up over 41.8\% in 2020 (Table~\ref{table:is_out}).

\subsection{Queer in AI Geography}

\begin{figure*}[hbtp]
\centering
\includegraphics[scale=0.3]{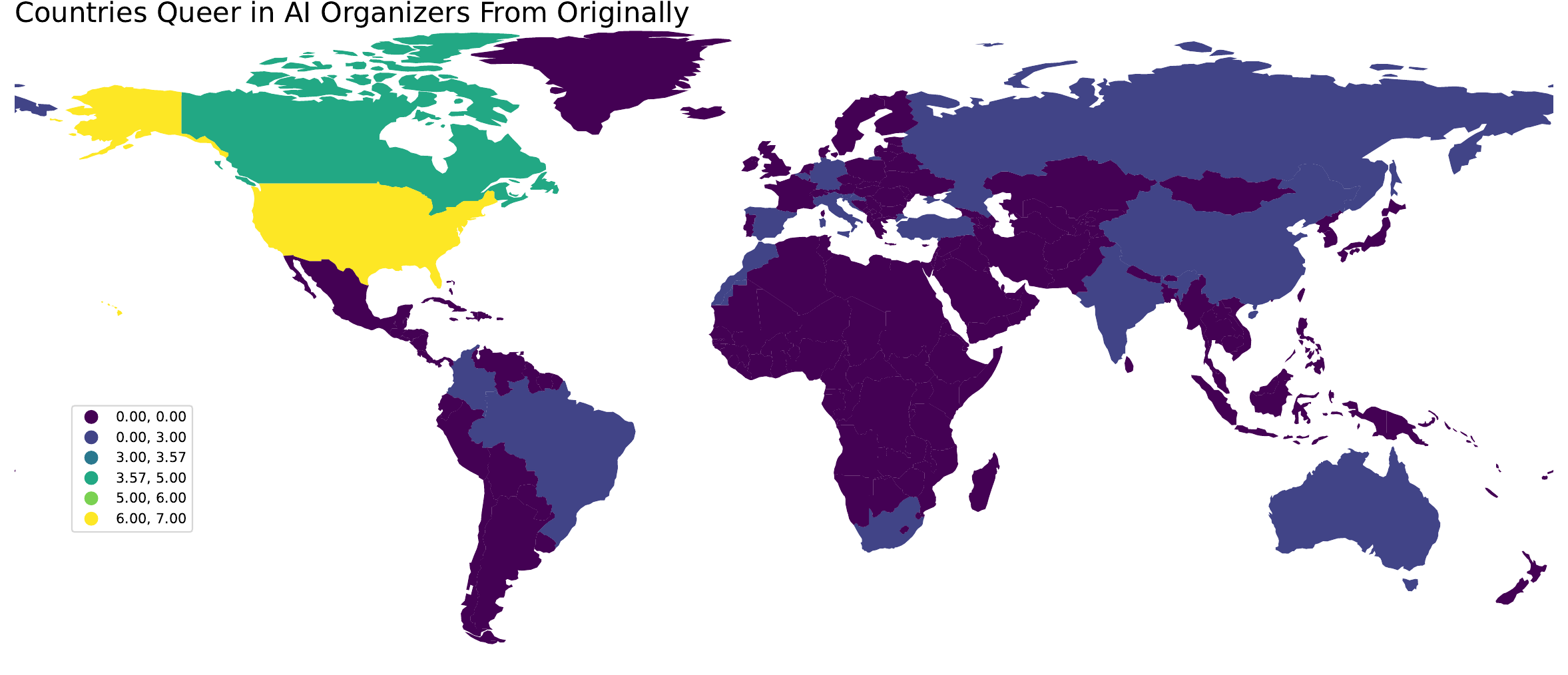}
\caption{Queer in AI Organizers Country of Origin.}
\label{fig:org_co}
\end{figure*}

\begin{figure*}[hbtp]
\centering
\includegraphics[scale=0.3]{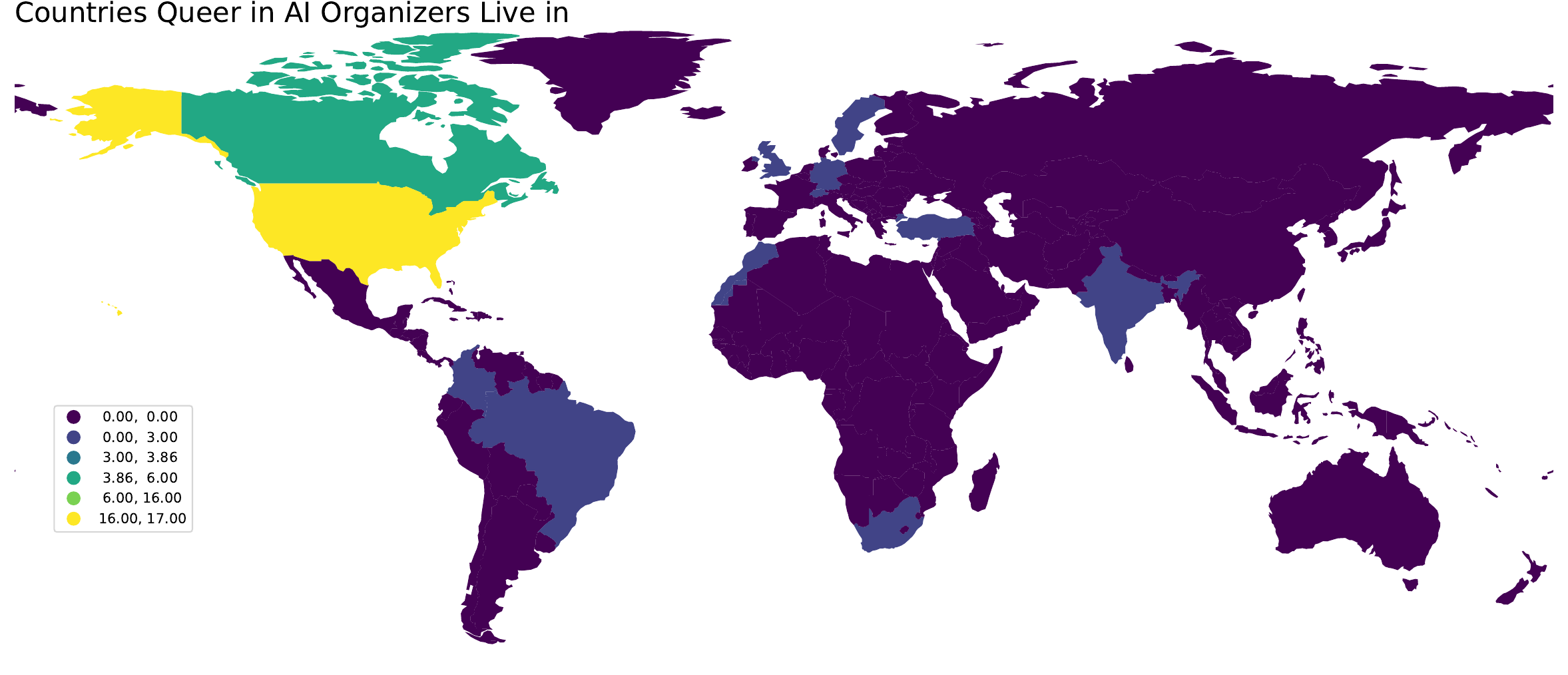}
\caption{Countries Queer in AI Organizers Live in.}
\label{fig:org_cl}
\end{figure*}

\begin{figure*}[hbtp]
     \centering
     \begin{subfigure}{1\textwidth}
         \centering
         \includegraphics[scale=0.3]{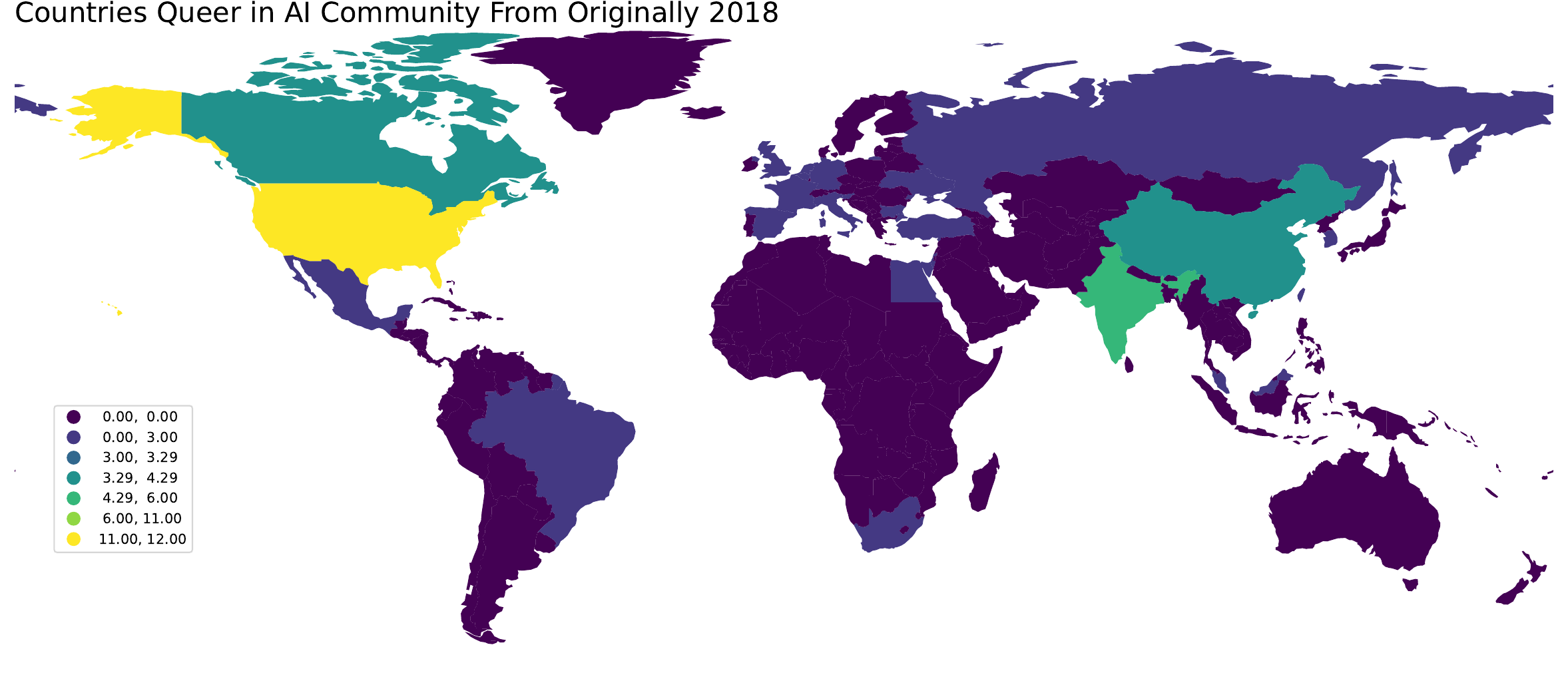}
         \label{fig:comm_co_18}
     \end{subfigure}

     \begin{subfigure}{1\textwidth}
         \centering
         \includegraphics[scale=0.3]{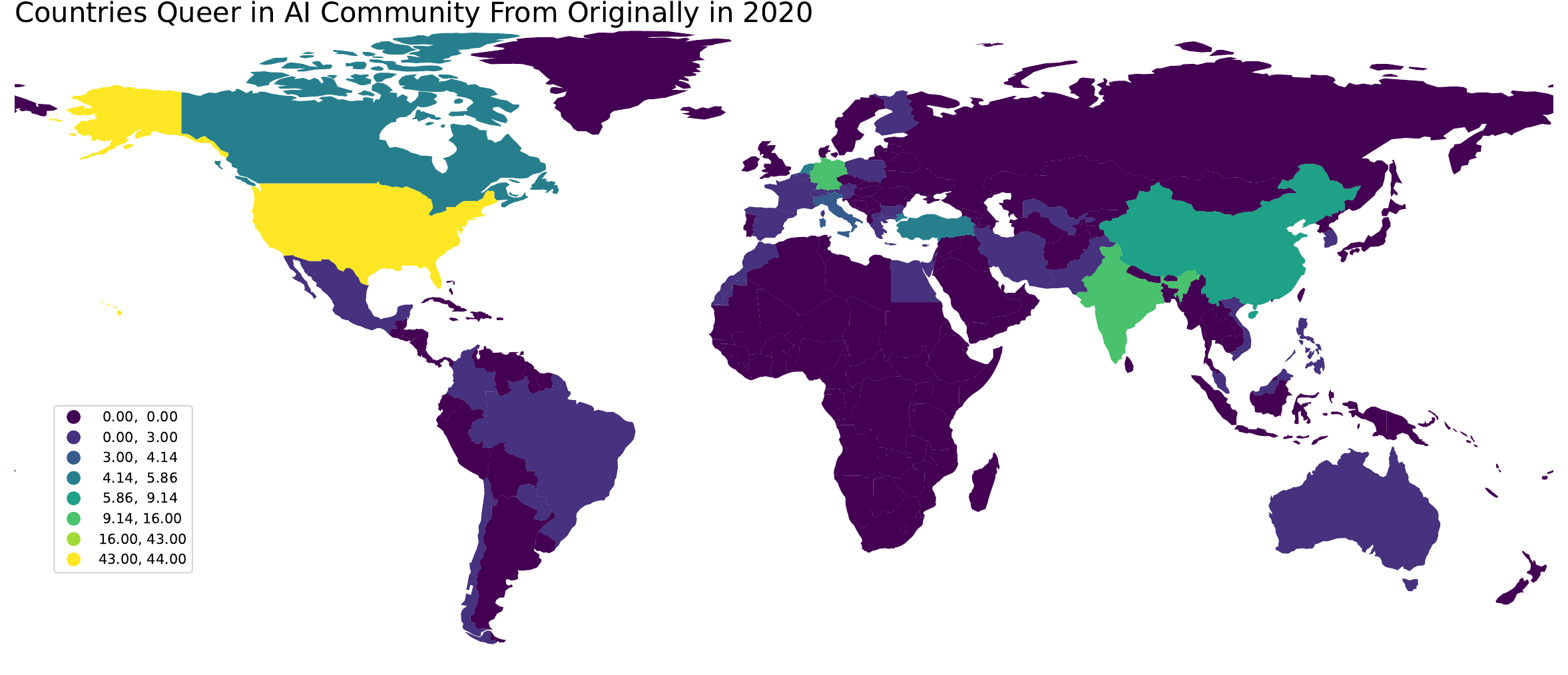}
         \label{fig:comm_co_20}
     \end{subfigure}
    
     \begin{subfigure}{1\textwidth}
         \centering
         \includegraphics[scale=0.3]{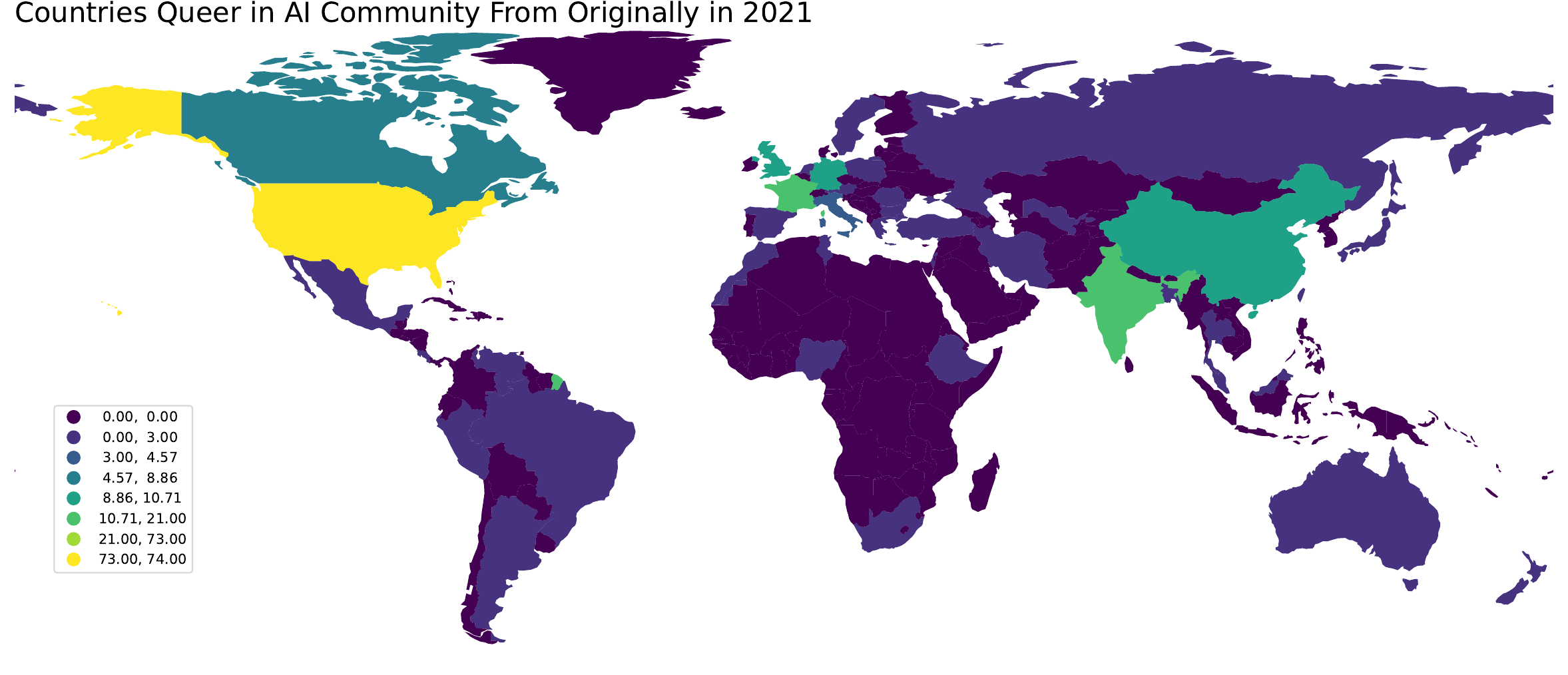}
         \label{fig:comm_co_21}
     \end{subfigure}
     \caption{Country Queer in AI Community From, 2018, 2020, 2021.}
     \label{fig:comm_co_all}
\end{figure*}

\begin{figure*}
     \centering
     \begin{subfigure}{1\textwidth}
         \centering
         \includegraphics[scale=0.3]{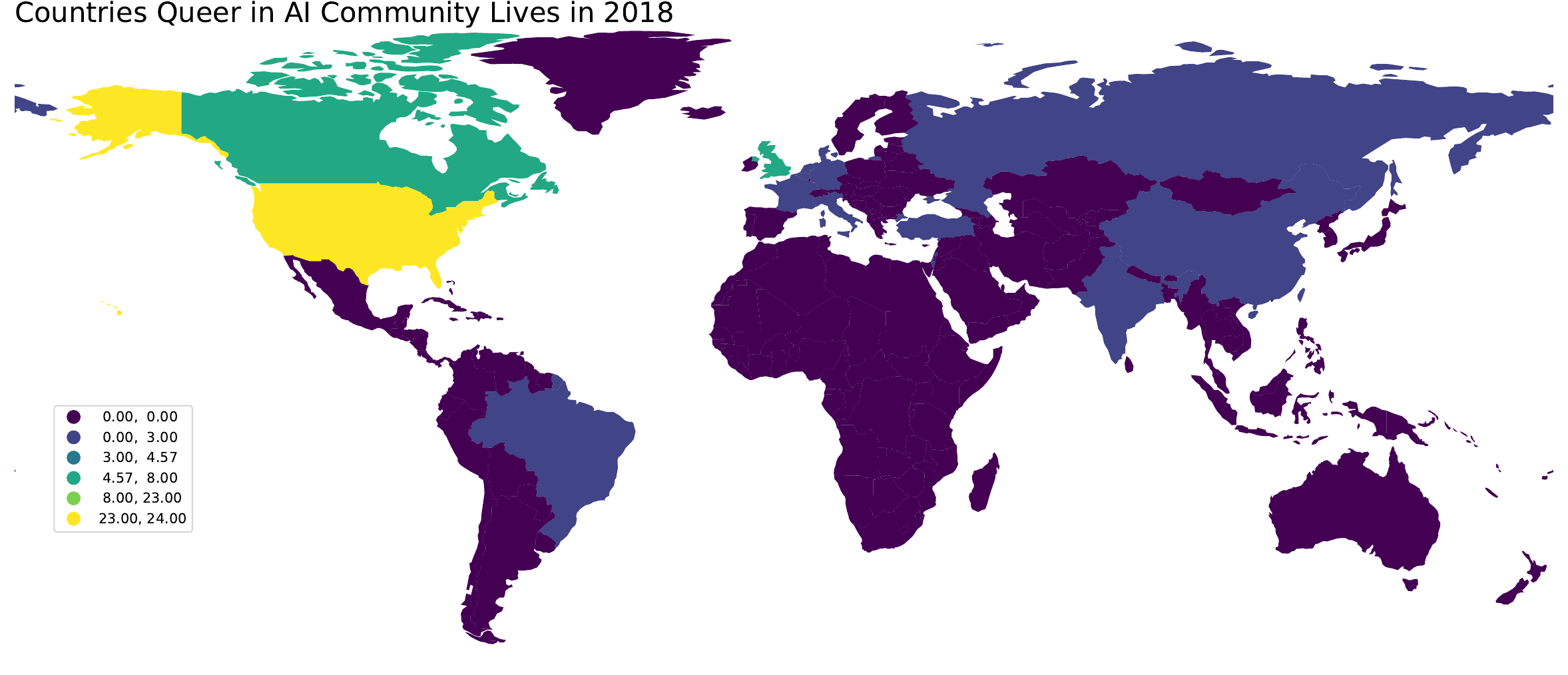}
         \label{fig:comm_cl_18}
     \end{subfigure}
     
     \begin{subfigure}{1\textwidth}
         \centering
         \includegraphics[scale=0.3]{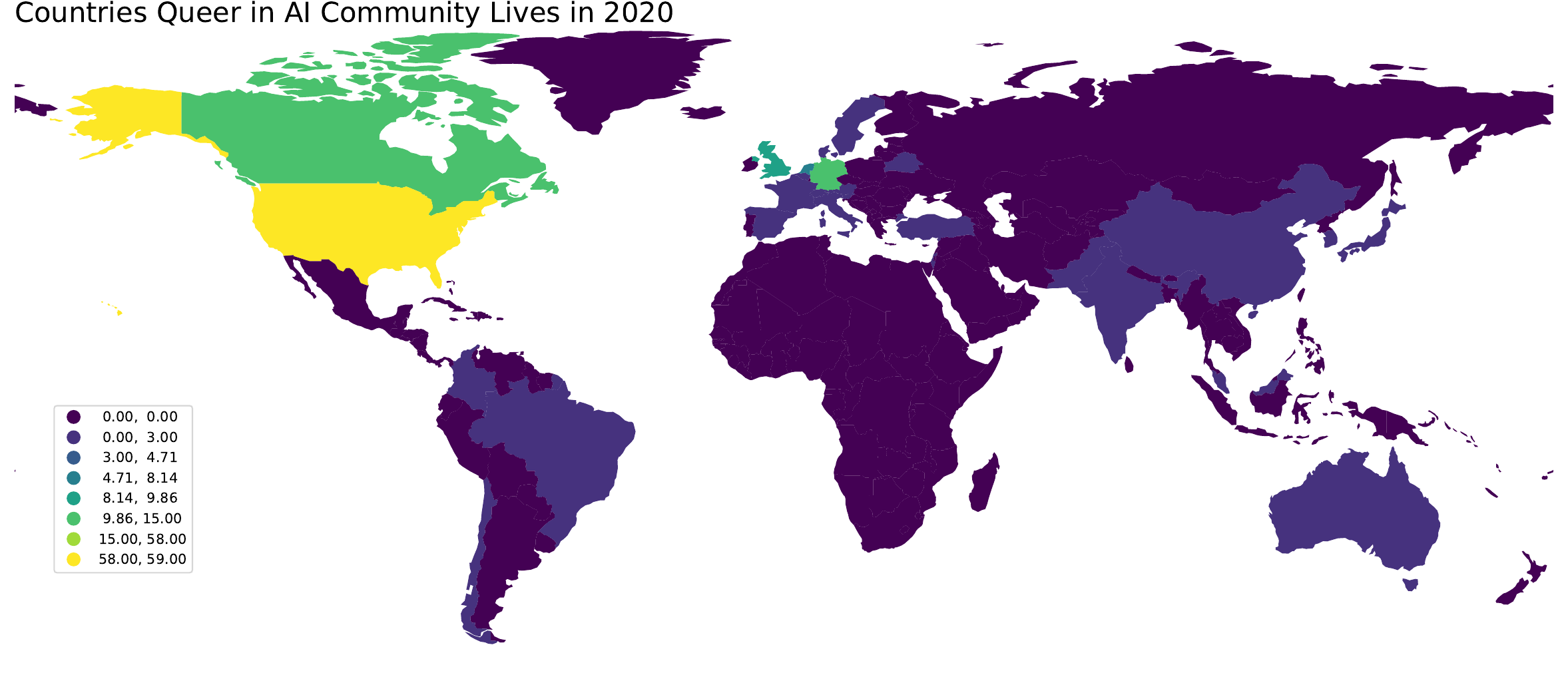}
         \label{fig:comm_cl_20}
     \end{subfigure}
    
     \begin{subfigure}{1\textwidth}
         \centering
         \includegraphics[scale=0.3]{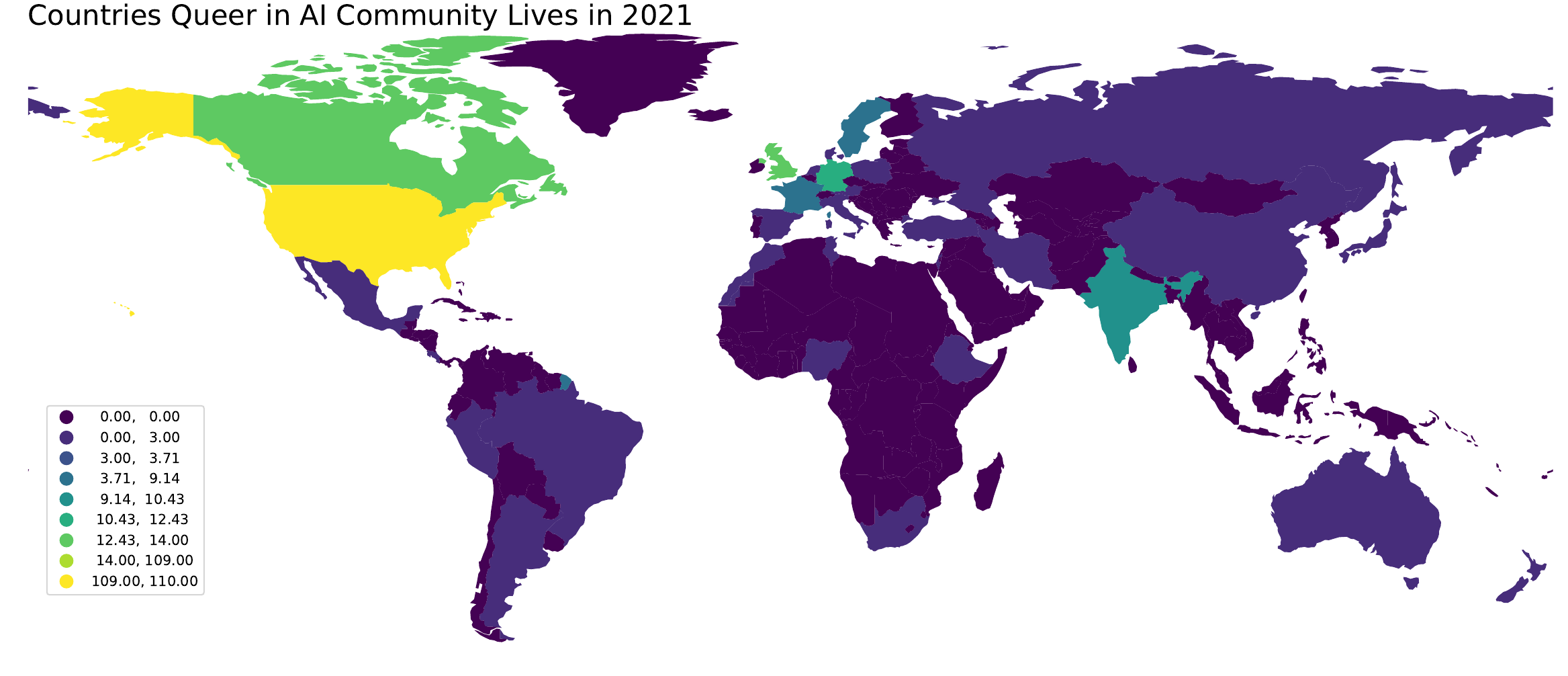}
         \label{fig:comm_cl_21}
     \end{subfigure}
     \caption{Country Queer in AI Community Lives in, 2018, 2020, 2021.}
     \label{fig:comm_cl_all}
\end{figure*}

\begin{figure*}[hbtp]
\centering
\includegraphics[scale=0.3]{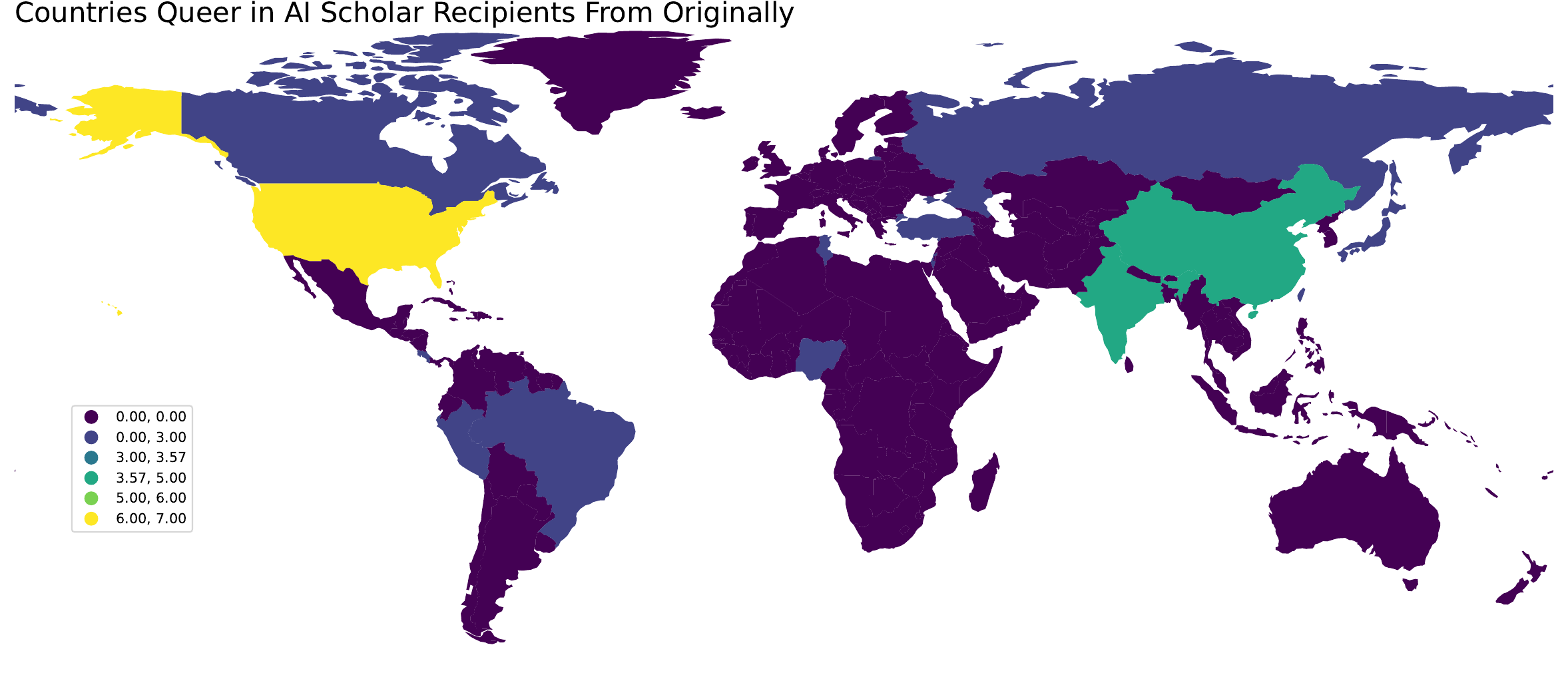}
\caption{Scholarship Recipients Country of Origin.}
\label{fig:sr_co}
\end{figure*}

\begin{figure*}
\centering
\includegraphics[scale=0.3]{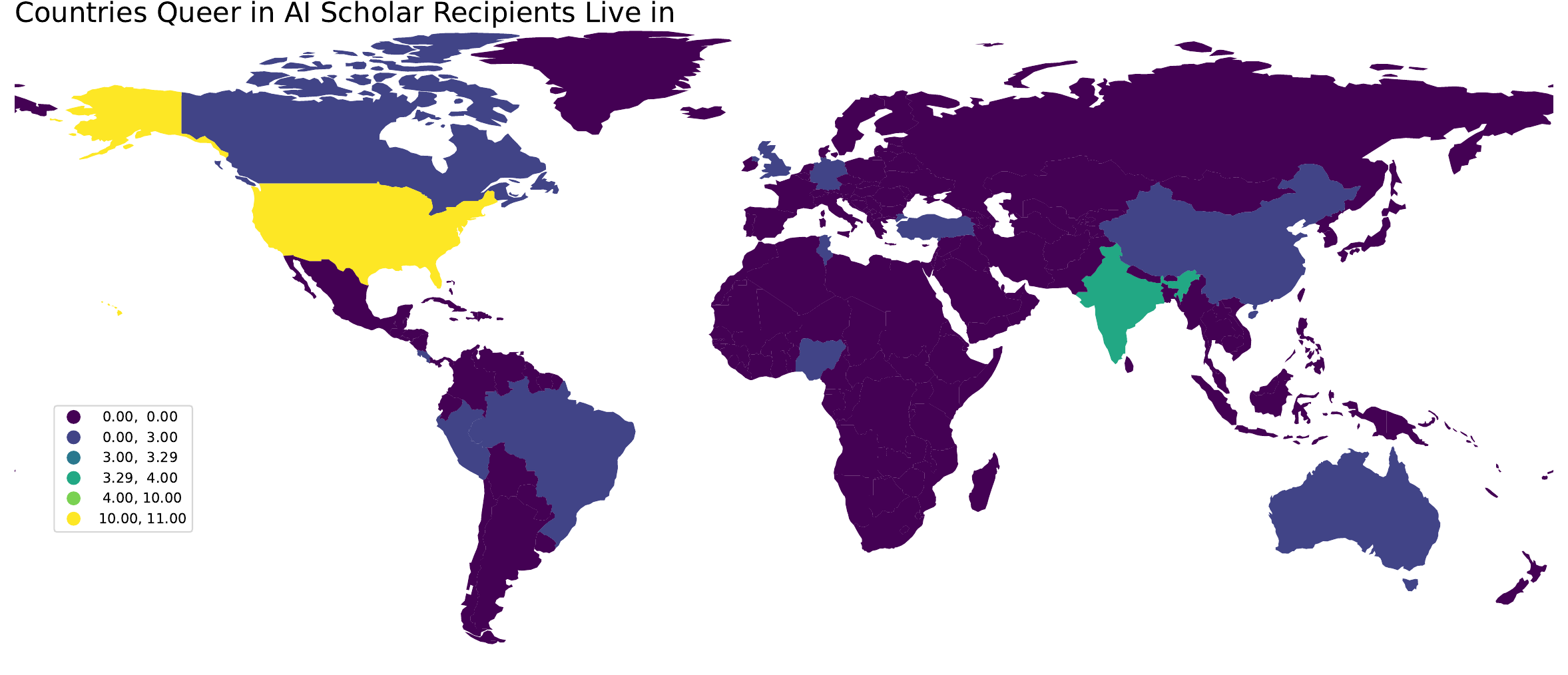}
\caption{Country Scholarship Recipients Live in.}
\label{fig:sr_ci}
\end{figure*}

This section reports the country of origin and residence for Queer in AI organizers, members, and graduate school scholarship recipients in Figures~\ref{fig:org_co}--\ref{fig:org_cl}, Figures~\ref{fig:comm_co_all}--\ref{fig:comm_cl_all}, and Figures~\ref{fig:sr_co}--\ref{fig:sr_ci} respectively.

\subsection{Reporting Survey Results: Ethnicity}
\label{ssec:aggregation}

Aggregating human data is complicated, and transparency in the process of aggregation is critical. We detail the choices that went into the aggregation of ethnicity responses. A group of organizers decided on a set of categories---``Mixed ethnicity,'' ``Black/African/African-American,'' ``Jewish,'' ``Southeast Asian,'' ``West Asian,'' ``East Asian,'' ``South Asian,'' ``Latinx,'' ``Caucasian,'' ``Middle Eastern,'' ``Hispanic,'' and ``Unaggregated''---after a pass of the raw data and getting feedback from multiple Queer in AI organizers and other affinity groups. The ``Unaggregated'' category was introduced to include responses that the organizers felt were not adequately described by the other existing categories (e.g. responses like `Person of color').

Aggregation for responses that belonged to two or more categories involved a few heuristics. Firstly, anything with the words `mixed' or `half' was assigned to the ``Mixed ethnicity'' category. Moreover, references to multiple categories were aggregated into each of the mentioned ones. For example, `half white and half Indian'\footnote{Example responses in this section are not taken from the raw data and provided for illustration only.} would be assigned to ``Mixed,'' ``Caucasian,'' and ``South Asian.'' For responses of the form `\{Ethnic identity\}--American', the organizers chose to consider only the mentioned ethnic identity, in alignment with the discussion among Queer in AI members. Given the complexity of the process, Queer in AI organizers continually evaluated best practices for including mixed and migrant ethnic identities in future surveys.

Although there is an overlap between the categories ``West Asian'' and ``Middle Eastern,'' organizers chose to retain both categories because ``Middle Eastern'' includes parts of Asia, Africa, and Europe. While the organizers recognize that ``Middle Eastern'' is a Eurocentric term, they chose to retain that as a category for this paper because several responses use that term. The organizers hope to improve this terminology in the future iterations after consulting the communities in question.

Lastly, a lot of responses self-described themselves as ``Asian.'' For each such response, the organizers looked at the corresponding response to country of origin, and used that to aggregate it within ``East/Southeast/West/Central/South Asian.'' Responses that did not clarify the country of origin were included within the ``Unaggregated'' category.

\begin{figure*}[t]
\centering
\subfloat[Queer in AI Community: Are you Trans?] {
    \includegraphics{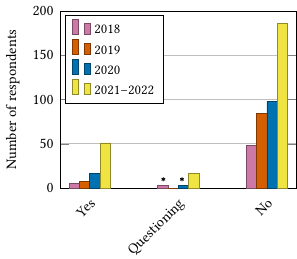}
    \label{fig:org_trans_c}
}\hspace{0.5cm}
\subfloat[Queer in AI Organizers: Are you Trans?] {
    \includegraphics{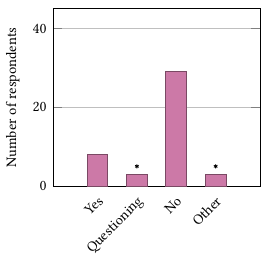}
    \label{fig:org_trans_o}
}\hspace{0.5cm}
\subfloat[Queer in AI Scholarship Recipients: Are you Trans?] {
    \includegraphics{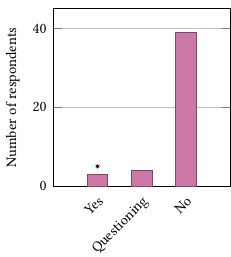}
    \label{fig:org_trans_sr}
  }
\caption{Queer in AI's statistics for transgender community members (\subref{fig:org_trans_c}), organizers (\subref{fig:org_trans_o}), and scholarship recipients (\subref{fig:org_trans_sr}) respectively. Write-in responses were aggregated by a team of Queer in AI organizers. For categories with $\leq3$ responses (marked by *), exact numbers are omitted to protect anonymity. \label{fig:trans}}
\end{figure*}

\begin{figure*}[t]
\centering
\subfloat[Queer in AI Community (2021--2022): Are you Intersex?] {
    \includegraphics{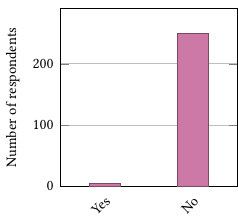}
    \label{fig:org_intersex_c}
}\hspace{1cm}
\subfloat[Queer in AI Organizers: Are you Intersex?] {
    \includegraphics{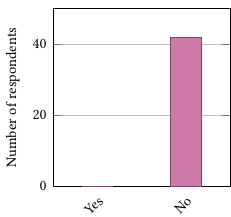}
    \label{fig:org_intersex_o}
}\hspace{1cm}
\subfloat[Queer in AI Scholarship Recipients: Are you Intersex?] {
    \includegraphics{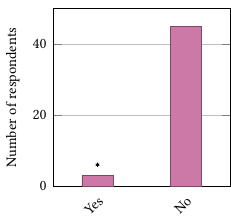}
    \label{fig:org_intersex_sr}
  }
\caption{Queer in AI's statistics for intersex community members (\subref{fig:org_intersex_c}), organizers (\subref{fig:org_intersex_o}), and scholarship recipients (\subref{fig:org_intersex_sr}) respectively. For categories with $\leq3$ responses (marked by *), exact numbers are omitted to protect anonymity. \label{fig:intersex}}
\end{figure*}

\begin{figure*}[t]
\centering
\subfloat[Queer in AI Community (2021--2022): Are you Neurodivergent or Disabled?] {
    \includegraphics{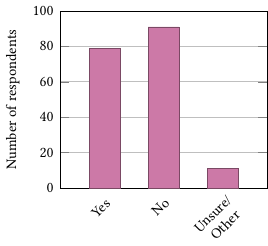}
    \label{fig:nd_comm}
}\hspace{1cm}
\subfloat[Queer in AI Organizers: Are you Neurodivergent or Disabled?] {
    \includegraphics{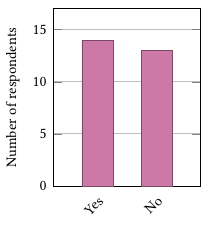}
    \label{fig:nd_org}
}\hspace{1cm}
\subfloat[Queer in AI Scholarship Recipients: Are you Neurodivergent or Disabled?] {
    \includegraphics{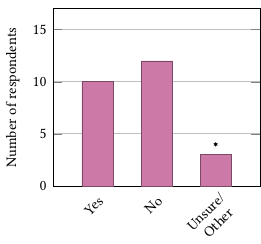}
    \label{fig:nd_scholar}
  }
\caption{Queer in AI's statistics for neurodivergent and/or disabled community members (\subref{fig:nd_comm}), organizers (\subref{fig:nd_org}), and scholarship recipients (\subref{fig:nd_scholar}) respectively. Write-in responses were aggregated by a team of Queer in AI organizers. \label{fig:nd}}
\end{figure*}

\begin{figure*}[t]
\centering
\includegraphics{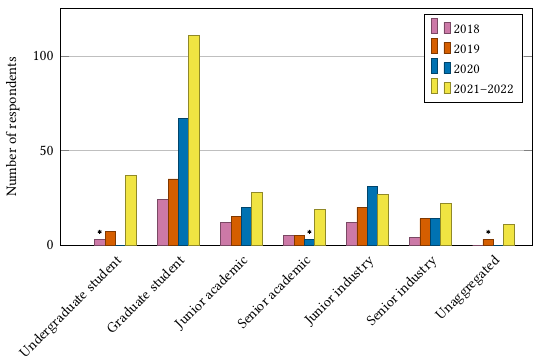}
\caption{Career stage statistics of the Queer in AI community members. Data was collected via the demographic surveys (\S\ref{ssec:demosurvey}). Write-in responses were aggregated by a team of Queer in AI organizers. ``Unaggregated’' refers to responses that could not be adequately described with any subset of other categories; however, responses in this group may overlap with the remaining categories. For categories with $\leq3$ responses (marked by *), exact numbers are omitted to protect anonymity.\\ 
\emph{Note: the 2020 survey did not differentiate between the graduate and undergraduate student respondents; here we report the combined total of these two groups under ``Graduate student''.} \label{fig:comm_career}}
\end{figure*}

\begin{table*}[hbtp]
\centering
\small
\begin{tabular}{lcccc}
\toprule
 & \textbf{Overall} & \begin{tabular}[c]{@{}c@{}}\textbf{Only non-Cis }\\\textbf{Respondents}\end{tabular} & \begin{tabular}[c]{@{}c@{}}\textbf{Only BIPOC }\\\textbf{Respondents}\end{tabular} & \begin{tabular}[c]{@{}c@{}}\textbf{Respondents who do not}\\\textbf{live in queer-safe countries}\end{tabular} \\
\midrule
\textit{I feared for my physical safety} & 23.6\% & 20.9\% & 26.5\% & 37.3\% \\
\textit{I felt I was deliberately ignored or excluded} & 43.6\% & 38.4\% & 49.0\% & 56.9\% \\
\textit{I felt intimidated/bullied} & 31.4\% & 27.9\% & 32.7\% & 43.1\% \\
\textit{I observed others staring at me} & 39.3\% & 31.4\% & 46.9\% & 39.2\% \\
\textit{I received a low performance evaluation} & 5.0\% & 4.7\% & 6.1\% & 7.8\% \\
\textit{I received unsolicited physical contact} & 22.9\% & 24.4\% & 20.4\% & 27.5\% \\
\textit{I was in a hostile work environment} & 20.0\% & 16.3\% & 20.4\% & 29.4\% \\
\textit{I was pursued, followed or stalked} & 10.7\% & 11.6\% & 12.2\% & 17.6\% \\
\begin{tabular}[c]{@{}l@{}}\textit{I was singled out as the "resident authority" }\\\textit{due to my identity}\end{tabular} & 42.1\% & 38.4\% & 32.7\% & 51.0\% \\
\begin{tabular}[c]{@{}l@{}}\textit{I was the target of derogatory comments }\\\textit{(written or in person)}\end{tabular} & 40.7\% & 39.5\% & 36.7\% & 49.0\% \\
\textit{I was the target of innuendos and/or jokes} & 47.9\% & 48.8\% & 49.0\% & 68.6\% \\
\textit{I was the target of vandalism or graffiti} & 3.6\% & 5.8\% & 4.1\% & 3.9\% \\
\bottomrule
\end{tabular}
\caption{Share of safety incidents reported by Queer in AI members in 2021.}
\label{table:safety}
\end{table*}

\begin{table*}[hbtp]
\centering
\small
\begin{tabular}{lcccc}
\toprule
 & \textbf{Overall} & \begin{tabular}[c]{@{}c@{}}\textbf{Only non-Cis }\\\textbf{Respondents}\end{tabular} & \begin{tabular}[c]{@{}c@{}}\textbf{Only BIPOC }\\\textbf{Respondents}\end{tabular} & \begin{tabular}[c]{@{}c@{}}\textbf{Neurodivergent}\\\textbf{Respondents}\end{tabular} \\
\midrule
\begin{tabular}[c]{@{}l@{}}\textit{Conferences exacerbate my mental }\\\textit{health problems~}\end{tabular} & 29.5\% & 28.2\% & 16.1\% & 29.0\% \\
\textit{I've had mental health crises at conferences~} & 16.8\% & 16.5\% & 21.0\% & 19.4\% \\
\textit{I've harmed myself~} & 30.2\% & 25.9\% & 25.8\% & 41.9\% \\
\begin{tabular}[c]{@{}l@{}}\textit{I've seriously considered~~or~}\\\textit{attempted suicide~}\end{tabular} & 38.3\% & 30.6\% & 38.7\% & 56.5\% \\
\begin{tabular}[c]{@{}l@{}}\textit{My ability to conduct research or participate}\\\textit{in classes has been impaired by mental~}\\\textit{health issues~}\end{tabular} & 79.9\% & 81.2\% & 58.1\% & 91.9\% \\
\bottomrule
\end{tabular}
\caption{Self-reported mental health of Queer in AI members in 2021.}
\label{table:mental_health}
\end{table*}

\begin{table*}[hbtp]
\centering
\small
\begin{tabular}{p{6cm}c}
\toprule
\multicolumn{2}{l}{\textbf{When did you feel your mental health was at the lowest point?}} \\
\midrule
\textit{When I was a student} & 43.2\% \\
\textit{Within the last year~} & 33.7\% \\
\textit{When I was questioning my sexual orientation / gender} & 31.1\% \\
\textit{When I was struggling with my career} & 27.4\% \\
\textit{Before and during coming out stages} & 26.8\% \\
\begin{tabular}[c]{@{}l@{}}\textit{When I was in the country where~}\\\textit{queer folks are generally not accepted}\end{tabular} & 15.3\%\\
\bottomrule
\end{tabular}
\caption{Moments when Queer in AI members have struggled the most with mental health.}
\label{table:mental_health_2}
\end{table*}

\begin{table*}[hbtp]
\centering
\begin{tabular}{lll}
\toprule
\begin{tabular}[c]{@{}l@{}}\textbf{Have you disclosed your gender and/or~}\\\textbf{sexual orientation to your peers?}\end{tabular} & \textbf{2020 Survey} & \textbf{2021 Survey} \\
\midrule
Completely out & 41.8\% & 47.7\% \\
Out except for certain people/not publicly & 22.4\% & 19.7\% \\
Out to friends/family & 6.1\% & 3.8\% \\
Out to certain friends/family & 27.6\% & 26.5\% \\
Not out to anyone & 2.0\% & 2.3\% \\
\bottomrule
\end{tabular}
\caption{Status of disclosure of gender and/or sexual orientation among Queer in AI survey respondents in 2020 and 2021.}
\label{table:is_out}
\end{table*}

\begin{table*}[hbtp]
\centering
\begin{tabular}{ll}
\toprule
\multicolumn{2}{l}{\textbf{On a scale of 1 to 5...}} \\
\midrule
\textit{...how well represented are non-cis folks in your immediate work/research group?} & 1.54 \\
\textit{...how well represented are non-cis folks in Queer spaces you are part of (this includes Queer in AI)?} & 3.12 \\
\textit{...how often have you witnessed transphobia / microaggressions against non-cis folks?} & 2.90 \\
\textit{...how often have you called out transphobia / microaggressions against non-cis folks?} & 2.90 \\
\textit{...how much effort have you put in to make your group more inclusive of non-cis folks?} & 3.10 \\
\textit{...how aware are you about the name change policies in academic publications?} & 2.67 \\
\textit{...how aware are you about issues regarding mitigating deadnaming in citations?} & 2.51 \\
\bottomrule
\end{tabular}
\caption{Average responses to survey questions from \textbf{cisgender respondents} about their familiarity with non-cisgender issues.}
\label{table:cis_fam}
\end{table*}

\begin{table*}[hbtp]
\centering
\begin{tabular}{ll}
\toprule
\multicolumn{2}{l}{\textbf{On a scale of 1 to 5...}} \\
\midrule
\textit{...how well represented are non-cis folks in your immediate work/research group?} & 1.98 \\
\textit{...how well represented are non-cis folks in Queer spaces you are part of (this includes Queer in AI)?} & 3.42 \\
\textit{...how often have you faced transphobia / microaggressions from cis folks?} & 3.01 \\
\textit{...how often have you faced transphobia / microaggressions from queer cis folks?} & 2.25 \\
\bottomrule
\end{tabular}
\caption{Average responses to survey questions from \textbf{non-cisgender respondents} about their experience with representation and transphobia.}
\label{table:noncis_fam}
\end{table*}

\begin{table*}[hbtp]
\centering
\begin{tabular}{ll}
\toprule
\multicolumn{2}{l}{\textbf{On a scale of 1 to 5...}} \\
\midrule
\textit{...how well represented are queer BIPOC folks in your immediate work/research group?} & 1.76 \\
\textit{...how well represented are queer BIPOC folks in Queer spaces you are part of (this includes Queer in AI)?} & 2.80 \\
\textit{...how often have you witnessed racism / microaggressions against queer BIPOC folks?} & 2.53 \\
\textit{...how often have you called out racism / microaggressions against queer BIPOC folks?} & 2.41 \\
\textit{...how much effort have you put in to make your group more inclusive of queer BIPOC folks?} & 2.91 \\
\bottomrule
\end{tabular}
\caption{Average responses to survey questions from \textbf{white respondents} about their familiarity with BIPOC issues.}
\label{table:white_fam}
\end{table*}

\begin{table*}[hbtp]
\centering
\begin{tabular}{ll}
\toprule
\multicolumn{2}{l}{\textbf{On a scale of 1 to 5...}} \\
\midrule
\textit{...how well represented are queer BIPOC folks in your immediate work/research group?} & 1.76 \\
\textit{...how well represented are queer BIPOC folks in Queer spaces you are part of (this includes Queer in AI)?} & 2.80 \\
\textit{...how often have you witnessed racism / microaggressions against queer BIPOC folks?} & 2.53 \\
\textit{...how often have you called out racism / microaggressions against queer BIPOC folks?} & 2.41 \\
\textit{...how much effort have you put in to make your group more inclusive of queer BIPOC folks?} & 2.91 \\
\bottomrule
\end{tabular}
\caption{Average responses to survey questions from \textbf{BIPOC respondents} about their experience with representation and racism.}
\label{table:bipoc_fam}
\end{table*}

\clearpage
\subsection{Finances} \label{app:finances}

Table~\ref{table:revenue} provides revenue sources for Queer in AI since its founding in 2018. Table~\ref{table:expenses} provides expenses of Queer in AI since 2018.

\definecolor{light-gray}{gray}{0.94}
\begin{table*}[hbt!]
\footnotesize
\centering
    \begin{tabular*}{\linewidth}{@{\extracolsep{\fill}} ccccc }
     \toprule
     \multicolumn{5}{c}{\cellcolor{light-gray} \bf 2018 Income \vspace{0.1cm}}\\
     Google & Twitter & Prowler.io & DeepMind & NVIDIA \\ 
      \$7,500 & \$2,500 & \$2,500 & \$2,500 & \$2,500 \\ 
    \midrule
    \end{tabular*}
    
    \begin{tabular*}{\linewidth}{@{\extracolsep{\fill}} cccccc }
     \multicolumn{6}{c}{\cellcolor{light-gray} \bf 2019 Income \vspace{0.1cm}}\\
      2018 carryover & Google & DeepMind & Microsoft & Prowler.io & Apple \\ 
     \$2,421.00 & \$20,000 & \$20,000 & \$20,000 & \$7,500 & \$15,000 \\ 
     \midrule
    \end{tabular*}
    
    \begin{tabular*}{\linewidth}{@{\extracolsep{\fill}} cccccccc }
     \multicolumn{8}{c}{\cellcolor{light-gray} \bf 2020 Income \vspace{0.1cm}}\\
     2019 carryover & Google & DeepMind & Microsoft & Apple & NVIDIA & Prowler.io & oSTEM\\ 
     \$56,884.52 & \$20,000 & \$20,000 & \$20,000 & \$20,000 & \$20,000 & \$7,500 & \$10,000 \\
     \midrule
    \end{tabular*}
    
    \begin{tabular*}{\linewidth}{@{\extracolsep{\fill}} cccccccccc }
    \multicolumn{10}{c}{\cellcolor{light-gray} \bf 2021 Income \vspace{0.1cm}}\\
     2020 carryover & Donations & DeepMind & Microsoft & Apple & oSTEM & NVIDIA & Capital One & Intel & Rasa Technologies \\ 
     \$130,852.30 & \$33,287.79 & \$20,000 & \$20,000 & \$20,000 & \$20,000 & \$15,000 & \$7,500 & \$7,500 & \$7,500 \\
     \midrule
    \end{tabular*}

    \begin{tabular*}{\linewidth}{@{\extracolsep{\fill}} ccccccccc }
    \multicolumn{9}{c}{\cellcolor{light-gray} \bf 2022 Income \vspace{0.1cm}}\\
     2021 carryover & Donations & Grants & DeepMind & Microsoft & Apple & oSTEM & D.E. Shaw Research & Netflix \\ 
     \$174,020.78 & \$13,710.78 & \$5,000 & \$20,000 & \$20,000 & \$20,000 & \$10,000 & \$15,000 & \$3,000  \\
     \bottomrule
    \end{tabular*}
    
\caption{Queer in AI revenue, in USD.}
\label{table:revenue}
\end{table*}

\begin{table*}[hbt!]
\centering
\begin{tabular}{ lrrrrr } 
     \toprule
     Expense & 2018 & 2019 & 2020 & 2021 & 2022 \\
     \midrule
     Physical Venue Costs & \$12,251.26 & \$22,018.02 & \$2,389.05 & -- & \$8,198.34\\
     Travel Scholarships and Registration & \$750.00 & \$2,053.20 & -- & \$200.00 & \$6,941.43 \\
     Promotional Goods or Items & \$2,065.00 & \$3,965.26 & \$426.48 & -- & \$55.43\\
     Honoraria for Speakers and Panelists & -- & -- & \$8,500.00 & \$20,000.00 & \$14,500.00\\
     Emergency Aid Program & -- & -- & \$10,000.00 & \$10,000.00 & \$5,000.00\\
     Consulting & -- & -- & -- & \$2,813.44 & --\\
     Virtual Infrastructure & \$12.00 & \$12.00 & \$90.00 & \$2,478.00 & -- \\
     Grad App Program & -- & -- & \$17,969.50 & \$71,880.23 & \$40,435.42\\
     Other & -- & -- & \$1,000.00 & -- & --\\
     Contractors & -- & -- & \$1,000.00 & -- & \$33,220.00\\
     \midrule
     Total & \$15,079.26 & \$28,036.48 & \$33,532.22 & \$107,371.67 & \$100,657.69 \\
     \bottomrule
    \end{tabular}
\caption{Queer in AI expenses, in USD.}
\label{table:expenses}
\end{table*}

\section{Policy, Advocacy, and Impact}
\label{sec:policy}

Queer in AI translates findings from its programs and the policies that it develops
into real-world advocacy and systemic change for queer researchers and communities.
Queer in AI, grounded in survey results and the inclusive conference guide,
has helped conferences such as
ICML \cite{icml}, NeurIPS \cite{neurips}, NAACL \cite{naacl}, EMNLP \cite{emnlp},
and more extensively revise their registration forms and diversity surveys to be more queer-inclusive;
helped shape author guidelines about publication accessibility, quality, and inclusivity;
worked actively with conference infrastructure and logistics teams to protect the privacy of queer speakers and participants;
helped institute effective name-change processes at NAACL and EMNLP;
and has worked with the Association for Computational Linguistics \cite{aclweb} to implement a name change process, proactive measures to prevent the deadnaming of trans authors, and protocols to handle authors' requests to keep their videos private.
These processes have at times been arduous,
encountering resistance due to bureaucratic inertia and lack of concern,
sometimes requiring substantial organizer effort and even threats of pulling Queer in AI events to effect.

Queer in AI also educates companies, universities, and the general public on queer inclusivity and the important, diverse topics discussed at Queer in AI workshops and socials.
Queer in AI has helped shape Semantic Scholar's feature to allow authors to indicate their pronouns on their profile, shared ways to improve queer inclusivity at venues including the Allen Institute for Artificial Intelligence, Nike, and the Toronto Public Library, and trained professors and teaching assistants at the University of California, Los Angeles on respecting and including queer students in their classrooms \cite{semanticscholar, allenai, nike, tpl}.
Resources built by Queer in AI related to the conference guide are also being adopted as exemplars by others, e.g.,\ at student ACM chapters.
Queer in AI also accessibly communicate the main points of its talks and panels through Twitter threads and maintains a public YouTube channel with talk and panel recordings \cite{qinaiyoutube}.

Queer in AI has also leveraged its findings and experience to push institutions for change.
They wrote to the National Science Foundation (NSF), criticizing the institution for failing to include LGBTQIA+ people in their diversity mission, study the discrimination and underrepresentation of LGBTQIA+ people, and even simply include questions on sexual orientation and gender identity in their STEM census surveys, despite Queer in AI's demographic survey results showing that queer scientists don't feel comfortable or welcome at conferences or work environments and face harassment and discrimination \cite{nsf}.
Queer in AI offered its expertise, drawn from the inclusive conference guide and years of experience running demographic surveys, to the NSF to implement questions on sexual orientation and gender identity while protecting the privacy of LGBTQIA+ data. 

\end{document}